\documentclass[12pt,showpacs,preprintnumbers,floatfix]{article}
\pdfoutput=1
\usepackage{graphicx} 
\usepackage{dcolumn} 
\usepackage{bm} 
\usepackage{amsfonts}
\usepackage{amsthm}
\usepackage{amsmath}
\usepackage{amssymb}
\usepackage{epsfig}
\usepackage{color}
\usepackage{textcomp}
\usepackage{hyperref}
\usepackage{subfigure}
\usepackage[utf8]{inputenc}
\usepackage[english]{babel}
\usepackage[usenames,dvipsnames,table]{xcolor}
\usepackage{enumitem}
\usepackage{mathabx}

\def\beq{\begin{equation}}
\def\eeq{\end{equation}}
\def\bea{\begin{eqnarray}}
\def\eea{\end{eqnarray}}
\newcommand{\obs}{\text{obs}}
\def\vEW{v_\text{EW}}
\def\LUV{\Lambda_\text{UV}}
\def\LQCD{\Lambda_\text{QCD}}
\def\mU{\mathcal U}
\def\xcosmo{x_\text{cosmo}}
\def\xmicro{x_\text{micro}}
\def\xmacro{x_\text{macro}}
\def\xlocal{x_\text{local}}
\def\qdm{q_\text{dm}}

\begin{document}

\thispagestyle{empty}
\begin{flushright}
    September 2025
\end{flushright}
\vspace*{1.7cm}
\begin{center}
    {\Large \bf Multiverse Predictions for Habitability:\\ Fundamental Physics and Galactic Habitability }\\

    \vspace*{1.2cm} {\large McCullen Sandora\footnote{\tt
            mccullen@bmsis.org}}\\
    \vspace{.5cm} 
    {\it  $^{}$Blue Marble Space Institute of Science}\\
    {\it  Seattle, WA 98154, USA}
    
    \vspace{2cm} ABSTRACT
\end{center}
In the multiverse hypothesis, a range of universes exist with differing values of our physical constants. Here, we investigate how the probabilities of observing our values of these constants depend on the assumptions made about the theories governing particle physics and cosmology, along with habitability. The particle physics effects we consider include constraints on the Higgs vacuum expectation value from big bang nucleosynthesis and supernovae, grand unified theories (GUTs), and standard model stability. Cosmology effects we consider are different theories of dark matter and baryogenesis, and for galactic habitability effects we include star formation efficiency, stellar encounters, supernova explosions, and active galactic nuclei. We find the following to be disfavored in the multiverse scenario: flexible GUTs, pessimistic galactic disruption rates, some origin of life theories, and freeze-out dark matter with high energy baryogenesis. These predictions can be tested in future experiments to either confirm or rule out the multiverse.

\vfill \setcounter{page}{0} \setcounter{footnote}{0}
\newpage

\section{Introduction}

The prevailing consensus among cosmologists today is that the multiverse is not a scientific theory because it cannot make testable predictions. This is the tenth paper in a series that aims to reverse this viewpoint by explicitly demonstrating that concrete predictions can be made in the multiverse theory. Our underlying logic is that the multiverse must be able to account for the values of the physical constants we observe, which entails constructing probability distributions for observing different values of each constant. These depend sensitively on the assumptions we make about habitability, and so, in the multiverse context, our presence in this universe is compatible with some habitability conditions and incompatible with others. Though we currently lack the knowledge to determine which habitability conditions are correct, future advances will give us a much better picture of where and under what conditions life can exist in the universe. When these conditions are eventually determined, we may then compare our findings to the predictions the multiverse theory has made about these matters to determine if the predictions were true. Because there are many different dimensions to habitability, these function as quasi-independent predictions, allowing us to gain strong confidence on whether the multiverse is true or false.

Because these predictions rely on a detailed account of many different aspects of habitability, our previous works strayed far from the usual multiverse discourse of cosmology and particle physics. Since we were concerned primarily with the many-faceted aspects of habitability, we relied on some heuristic shortcuts when deriving our results, and neglected the impact the underlying theories of cosmology and particle physics have on the probabilities we compute. This paper remedies this by first explicitly presenting our formalism from a first principles approach. This allows us to track how our results depend on the assumptions we make about the theories governing the physical constants present in the standard models of cosmology and particle physics.

Our focus is on deriving probabilities for the six macroscopic dimensionless constants that most strongly dictate the structure of our universe. As we detail in the next subsection, these probabilities depend on the relative fraction of universes with each value of the constants, how habitability depends on these constants, and what we call the \emph{induced weight}, which acts as a posterior modification to the prior probability distribution for the constants. It is this induced weight that captures the dependence on the fundamental theories of cosmology and particle physics.

Secondly, we take this opportunity to augment our calculations to include a new macroscopic variable related to the galactic density, which was not included previously. This extension necessitates an inclusion of some galactic habitability considerations, which we discuss here. We show that these effects alter some of our previous conclusions, and act to more tightly constrain the range of habitability conditions which are compatible with the multiverse.

\subsection{Formalism}
In the multiverse framework, the constants of physics are not derivable from first principles, but instead vary from universe to universe. As such, the theory does not predict the exact constants we observe. However, we may still make sure that the theory is compatible with our observations by ensuring that the predicted probability of our observations is reasonably close to 1. Generically, we specify our observations as a set of parameters $\{x^\obs_i\}$, which include physical constants such as the masses of particles, strengths of forces, etc., and local variables, which include the mass of our star and planet, the timing of our species' occurrence, the water content of our planet, etc. The probability of observing our specific values of these parameters may then be written as a function of separate observables $\mathcal O(x)$ as 


\beq
\langle\mathcal O\rangle \,\propto \int dx\, p(x)\,\mathcal O(x)N_{\frac{\text{observers}}{\text{universe}}}(x)\label{lOr}
\eeq

Here, the observables may be functions like $\theta(x_i-x^\obs_i)$, where $\theta(x)$ is the Heaviside function, which specifies the cumulative probability of observing a value of a particular fundamental constant or local environmental variable at least as large as ours. Additional observables take the form $\theta(f(x_i)-f^\obs)$ for some function $f$, representing the probability of observing some derived variable at least as large as ours. Quantities of this latter type are useful, for example, when determining the probability of being in a universe with our carbon to oxygen ratio, as in \cite{mc5}. The quantity $ p(x)$ is a measure representing the relative frequency of universes with particular values of physical constants. This formalism also makes use of the principle of mediocrity \cite{mediocre}, whereby the probability of being in a particular universe is proportional to the number of observers within that universe. 

The physical constants can usefully be split into four categories: microscopic, macroscopic, and cosmological, and local. The macroscopic constants are those that directly dictate physical aspects of our universe, and are comprised of the fine structure constant $\alpha$, the masses of the electron $m_e$, proton $m_p$, up quark $m_u$ and down quark $m_d$ and more usefully their dimensionless mass ratios $\beta=m_e/m_p$, $\delta_u=m_u/m_p$, and $\delta_d=m_d/m_p$. Also included is the strength of gravity $\gamma=m_p/M_{pl}$, and the density parameter of galaxies $\kappa$, which is derived from cosmological variables in section \ref{cosmovar}. The microscopic parameters are those remaining parameters of the standard model of particle physics that do not directly influence the habitability properties of the macroscopic world. Similarly, the cosmological parameters are those of the standard model of cosmology that do not directly influence the macroscopic world.

The prior can be factored as 
\beq
 p(x)=
 p_\text{micro}(\xmicro,\xmacro)\,
 p_\text{cosmo}(\xcosmo,\xmacro)\,
 p_\text{local}(\xlocal,\xmacro)\,
\eeq
The various priors in this expression depend not only on the relevant subset of variables, but also the macroscopic constants, through both anthropic thresholds, as well as additional factors that are introduced when changing variables, to be discussed in sections \ref{microvar} and \ref{cosmovar} below.

The number of observers per universe can be split into a product of factors:

\beq
N_{\frac{\text{observers}}{\text{universe}}}(x) = N_{\frac{\text{protons}}{\text{universe}}}(\xcosmo,\xmacro)\,N_{\frac{\text{stars}}{\text{proton}}}(\xcosmo,\xmacro)\,
N_{\frac{\text{observers}}{\text{star}}}(\xmacro,\xlocal)
\eeq

The number of protons per universe will be discussed in section \ref{cosmovar}. The number of stars per proton can be written as $N_{\frac{\text{stars}}{\text{proton}}}(\xcosmo,\xmacro)=\epsilon_\text{SF}(\xcosmo,\xmacro)/\langle N_{\frac{\text{protons}}{\text{star}}}(\xmacro)\rangle$, where $\epsilon_\text{SF}$ is the star formation efficiency which depends on both cosmological and macroscopic constants, as will be discussed in section \ref{cosmovar}. The average number of protons per star depends only on macroscopic constants, and is given by $\langle N_{\frac{\text{protons}}{\text{star}}}(\xmacro)\rangle=27.7\alpha^{3/2}\beta^{-3/4}\gamma^{-3}$ \cite{mc1}. The number of observers per star depends on both macroscopic constants and local parameters, and is highly sensitive to assumptions made about habitability. This quantity has been the focus of our previous papers in this series, and can be referred to as stellar habitability $\mathbb{H}$.

With this factorization, the microscopic and cosmological variables can be isolated into induced weight functions for the relevant macroscopic variables $W_\text{micro}(\xmacro)$ and $W_\text{cosmo}(\xmacro)$, respectively, along the lines of \cite{feldstein2005density}.
\bea
\langle\mathcal O\rangle \,\propto \int d\xmacro\,d\xlocal\, W_\text{micro}(\xmacro)\, W_\text{cosmo}(\xmacro)\times\nonumber\\
\mathcal O(\xmacro,\xlocal)\,
\frac{ p_\text{local}(\xlocal,\xmacro)}{N_{\frac{\text{protons}}{\text{star}}}(\xmacro)}\,
\mathbb{H}(\xmacro,\xlocal)\label{eqnOWW}
\eea
where
\beq
W_\text{micro}(\xmacro)=\int d\xmicro\, p_\text{micro}(\xmicro,\xmacro)
\eeq

and
\beq
W_\text{cosmo}(\xmacro)=\int d\xcosmo\, p_\text{cosmo}(\xcosmo,\xmacro)\,
N_{\frac{\text{protons}}{\text{universe}}}(\xcosmo,\xmacro)\,
\epsilon_\text{SF}(\xcosmo,\xmacro)\label{Wcosmo}
\eeq

Lastly, we note that in this expression, we have restricted our attention to universes whose laws have the same form as our own, so that only the values of our physical constants are varied. Equation \ref{lOr} may in principle be extended to additionally sum over the full suite of possibilities, including universes with different particle content, forces, number of dimensions, etc., to arrive at even more constraining probabilities than we consider. However, at the current moment, we regard such calculations as both intractable and untestable, and so content ourselves with the more restricted analysis.

In section \ref{microvar} we derive the induced weight for the microscopic variables, including constraints on the Higgs vacuum expectation value, grand unified theories, and standard model stability. In section \ref{cosmovar} we derive the cosmological induced weight, including its dependence for several different theories of dark matter and baryogenesis. This section includes a discussion on the number of protons per universe and how galactic density may affect habitability for stellar encounters, supernova explosions, and active galactic nuclei. In section \ref{results} we discuss our results, including predictions for which fundamental theories and contributions to galactic habitability are compatible with our observations in the multiverse context.

\section{Microscopic Variables}\label{microvar}

The relevant microscopic variables of the standard model include the force strengths $\{\alpha_i\}$, the Yukawa couplings $\{\lambda_i\}$, the electroweak vacuum expectation value $\vEW$, and the Higgs self-coupling $\{\lambda_H\}$. We also include in this list the Planck mass $M_{pl}$ parameterizing the strength of gravity. Additional microscopic parameters are the CP-phase angles $\{\theta^\text{CP}_i\}$, neutrino masses $\{m^\nu_i\}$, and the parameters governing any axion sector. These additional parameters are presumed here to not strongly influence the structure of our macroscopic world, and as such factor out into an irrelevant prefactor in the measure, though many exceptions to this expectation have been proposed in the theoretical physics literature (for instance, the Higgs may couple to dark matter for stability reasons, introducing a correlation between dark matter abundance and Higgs mass \cite{hertzberg2017correlation}, there are some models of dark energy that strongly involve the neutrino sector \cite{niedermann2023new}, and axions may play a ubiquitous role in potentially every unknown sector of cosmology \cite{arvanitaki2010string}.)

Though the probability distributions governing each of these quantities may only be computed within the context of a fully complete theory of everything, we may make reasonable ansatzes for each of these variables to determine the measure $ p_\text{micro}$. In \cite{mc8} it was found that different reasonable choices of measure do exert some influence on the resulting probabilities, but the influence is not nearly as strong as the different choices of habitability conditions we consider. This measure is most easily phrased in terms of particle physics parameters initially, but once established we transform the measure to be written in terms of our macroscopic constants.

The standard model is an effective field theory which is only supposed to be valid up to some high energy scale $\LUV$. Though this scale could in principle be only slightly larger than the energies we've probed with our current most powerful particle accelerators, we take the scale $\LUV$ to be on the order of the Planck mass $M_{pl}$, representing the scale at which our current description of reality truly becomes invalid (particles may become strings, spacetime may cease to be a useful concept, etc.). The conclusions we arrive at in this article will not depend on this choice of treatment. We generically expect the electroweak vacuum expectation value $\vEW$ to be governed by the distribution $\vEW^2\sim\mU(0,\LUV^2)$, where $\mU$ is the uniform distribution (this is the \emph{hierarchy problem}, where we generically expect $\vEW$ to be set by high scale physics; see for example \cite{zee2010quantum}). Similarly, we expect every particle coupled to gravity (that is, every particle) to contribute to the Planck mass, so that $M_{pl}^2\sim\mU(0,\LUV^2)$. The force strengths $\{\alpha_i\}$ are assumed to be uniform, $\alpha_i\sim\mU(0,1)$, (though since these are equal to charge couplings squared, it may also reasonably be conjectured that these come from a distribution $p(\alpha_i)\sim1/\alpha_i^{1/2}$). Similarly, we adopt that the Higgs self coupling is drawn from $\lambda_H\sim\mU(0,1)$. In \cite{leptonland} it was argued that the Yukawa couplings are log-uniformly distributed, $\lambda_i\sim1/\lambda_i$, on the basis that the observed Yukawa couplings are consistent with this distribution, as well as theoretical expectations.
With these assumptions, the microscopic measure becomes
\beq
d\xmacro\,d\xmicro\,  p_\text{micro}(\xmicro,\xmacro) =  \frac{d\vEW^2\,dM_{pl}^2}{\LUV^4}\prod_id\alpha_i\prod_j\frac{d\lambda_j}{\lambda_j}d\lambda_H d\mu_\text{irrel}\label{mumicro}
\eeq
Where the factor $d\mu_\text{irrel}=dx_\text{irrel}\, p_\text{irrel}(x_\text{irrel})$ contains all the fundamental variables that play no role in the macroscopic world, such as the axion, CP sector, etc. discussed above.

There is an inherent ambiguity in how we have treated the force strengths up to this point, as these depend on the energy scale used to measure them. For the weakly coupled forces, electromagnetism and the weak nuclear force, we may simply use the asymptotic low energy values, $\alpha$ and $\alpha_w$. For the strong nuclear force, this instead may be replaced with the energy scale at which the force becomes strongly coupled, through the process of dimensional transmutation \cite{coleman1973radiative}. This gives an energy scale $\LQCD\sim\LUV \exp(-2\pi/(9\alpha_\text{QCD}(\LUV)))$ \cite{hall2014weak}\footnote{A more thorough analysis would replace $\LUV$ in this expression with $\LUV^{7/9}v^{2/9}(\lambda_t\lambda_b\lambda_c)^{2/27}$, but this does not affect the resulting measure.}. This QCD energy scale gives the dominant contribution to the proton mass, and so we may rewrite this expression in terms of the strength of gravity $\gamma$ as $\LQCD/\LUV\sim c m_p/M_{pl}\equiv c \gamma$. With this, we have $d\alpha_\text{QCD}= f(\gamma)d\gamma/\gamma$, where $f(\gamma)$ is some function that depends only logarithmically on $\gamma$. In our analysis we will treat $f(\gamma)$ as a constant because these subleading contributions would be inappropriately precise when coupled with the back of the envelope nature of our habitability conditions.

The Yukawa couplings are related to quark and lepton masses through $m_i=\lambda_i \vEW$. To make these dimensionless, we consider the ratio to the proton mass. Of particular importance is the electron to proton mass ratio, $\beta=m_e/m_p$. All other ratios will be denoted as $\delta_i=m_i/m_p$, with the exception of the top quark, for which we will find it to be more convenient to define $\Delta_t=m_t/M_{pl}$. Generically, in equation \ref{mumicro} we may replace $d\lambda_i/\lambda_i$ with $d\delta_i/\delta_i$. We also define the dimensionless ratio $\rho_\text{EW} = \vEW/M_{pl}$, so that $dv$ may be replaced with $M_{pl}d\rho_\text{EW}$. The Higgs self coupling may be replaced with the Higgs mass through the formula $m_H^2=2\lambda_H\vEW^2$ (see for example \cite{zee2010quantum}), and we may define the dimensionless ratio $\sigma_H=m_H/M_{pl}$, so that $d\lambda_H$ may be replaced with $\sigma_H d \sigma_H/\rho_\text{EW}^2$. Finally, we define the ratio $q_{pl}=M_{pl}/\LUV$. With these, the microscopic measure can be rewritten as 
\beq
d\xmacro\,d\xmicro\,  p_\text{micro} = q_{pl}^3dq_{pl}\,\frac{d\rho_\text{EW}}{\rho_\text{EW}}\,\sigma_H d\sigma_H\, d\alpha\, d\alpha_w\, \frac{d\gamma}{\gamma}\,\frac{d\beta}{\beta}\,\frac{d \delta_u}{\delta_u}\,\frac{d\delta_d}{\delta_d}\frac{d\Delta_t}{\Delta_t}\,d\mu_\text{irrel}
\eeq
Here, we have only explicitly expressed the dependence on the up, down and top quarks, absorbing the heavier quarks and leptons into the factor $d\mu_\text{irrel}$. This expression makes clear the hierarchy problem, wherein there is a strong pressure for the Higgs mass to be close to the high energy cutoff scale $\LUV$, rather than its observed value which is minuscule in comparison. There is an even stronger pressure for the Planck mass to be as large as possible, aligned with the enormous value we observe. No such pressure is present for the electroweak vacuum expectation value, however.

When the microscopic variables are integrated over, this leads to an induced weight for the relevant physical constants $\alpha$, $\beta$, $\gamma$, $\delta_u$ and $\delta_d$ \footnote{What we call the electron is really the lightest of the three leptons, and so could actually be governed by a different distribution $p_e\sim p c^2$, with $p$ and $c$ the probability and cumulative density functions of lepton masses. This correction is only logarithmic, however.}. In the most basic scenarios the integral over microscopic variables can be performed implicitly, as it has no dependence on any of these constants. However, in certain instances some constraints, either arising from the structure of the fundamental physical theory or through anthropic boundaries, may give rise to additional dependence on the relevant physical constants. In these cases an additional contribution to the induced weight $W_\text{micro}$ may result. Below we detail three scenarios that potentially alter the measure, split by the constants they involve: $\vEW$, which may be influenced by the production of elements during big bang nucleosynthesis or in type II supernovae, $\alpha_w$, which may be constrained in a grand unified theory, and the constants $\sigma_H$ and $\Delta_t$, which may be constrained by standard model vacuum stability. Generically, we have
\beq
W_\text{micro}(\xmacro)=W_\text{EW}(\gamma)\,W_\text{Ht}(\gamma)\,W_\text{weak}(\alpha,\gamma)
\eeq

\subsection{Higgs VEV}

The electroweak vacuum expectation value (VEV) can influence macroscopic properties of the universe in two ways: through the final abundance of elements during big bang nucleosynthesis (BBN), which is the dominant source of hydrogen and helium in our universe, and the physics of type II supernova explosions, which is the dominant source of medium weight elements, including the biologically relevant elements oxygen and phosphorus. We discuss each of these influences in turn, and the constraints these introduce on the allowable parameter space, if these effects are taken to be important for the development of life.

The first effect governs the resulting helium abundance from BBN. In our universe, this is $Y_4=4\text{He}/(4\text{He}+\text{H})=.25$, reflecting a partial but incomplete conversion of hydrogen to helium in the early universe. This is a direct result of the final neutron to proton ratio at the end of BBN, and the fact that this ratio is neither approximately 0 nor 1 is due to the fact that the neutron lifetime is comparable to the BBN timescale, which are both measured in minutes \cite{mukhanov2005physical}. In \cite{hall2014weak}, the condition for leftover hydrogen from BBN was written in terms of constants as $\vEW\lesssim(m_n-m_p)^{3/4}M_{pl}^{1/4}$. 

While the presence of helium immediately after BBN may not have any direct impact on life, the authors note that if all hydrogen were depleted in this process, the effects on life could be adverse for three reasons: halo cooling would take longer, hydrogen burning stars would be absent, and hydrogen would not be available for use in biochemistry. This consideration therefore places a one sided bound on the fundamental constants to ensure that hydrogen is present in the universe. If we use the difference between neutron and proton masses from \cite{hall2008evidence}, we can write this as 
\beq
\rho_\text{EW}\lesssim\left(\delta_d-\delta_u-.18\alpha\right)^{3/4}\gamma^{3/4}
\eeq
This induces a contribution to the measure
\bea
W_\text{EW}&=&\int\frac{d\rho_\text{EW}}{\rho_\text{EW}}\theta\left(\left(\delta_d-\delta_u-.18\alpha\right)^{3/4}\gamma^{3/4}-\rho_\text{EW}\right)\nonumber\\
&\propto&\text{const}+\frac34\log\bigg(\left(\delta_d-\delta_u-.18\alpha\right)\gamma\bigg)
\eea
Where here, $\text{const}=-\log\rho_\text{min}$. This expression only depends logarithmically on the constants, and so will be irrelevant for our analysis, again because it is subleading.

The second constraint on $\vEW$ governs the dynamics of core collapse (type II) supernovae, and is reliant on a coincidence of length scales. In these systems, sudden exhaustion of nuclear fuel removes the supporting pressure that had previously stabilized the stellar radius, resulting in a rapid compression of material. One proposed mechanism for how this collapse triggers an explosion relies heavily on neutrino physics; as matter falls inwards, high enough densities are reached to convert electrons into neutrinos, which through their weak cross section can escape the inner core. However, the cross section is not weak enough for these neutrinos to escape the star completely, resulting in the bulk of the energy they carry being imparted in the outer stellar envelope. This causes the outer material to be ejected from the system, providing heavy elements for the next generation of stars \cite{mirizzi2016supernova}. This balancing act relies on the coincidence that the neutrino mean free path $l_\nu\sim1/(n_\text{core}\sigma_\nu)\sim\vEW^4/m_p^5$ is comparable to the core radius, $R_\text{core}\sim (M_\star/\rho_\text{core})^{1/3}\sim M_{pl}/m_p^2$ \cite{d2019direct}. In terms of our dimensionless variables, this becomes $\rho_\text{EW}\sim\gamma^{3/4}$. There it was found through simulations and analytic arguments that if $\vEW$ were about 20$\%$ larger, neutrinos would stream right through the star's outer layers, leaving the rich deposits behind. If $\vEW$ were about 5 times smaller, the neutrinos would get caught in the dense core, and the system would collapse entirely into a black hole. This range is narrow enough that we may approximate this constraint as a delta function in these variables, resulting in the following contribution to the measure:
\beq
W_\text{EW}=\int \frac{d\rho_\text{EW}}{\rho_\text{EW}}\delta\left(\rho_\text{EW}-\gamma^{3/4}\right)=\frac{1}{\gamma^{3/4}}\label{WEWSN}
\eeq

There are three reasons this factor may be disregarded in the measure: 1) medium weight elements such as oxygen and phosphorus may not actually be essential for life, 2) additional sources of these elements, such as Wolf-Rayet and CNO stars, may be sufficient sources of these elements, and 3) neutrinos may turn out to not actually be necessary to trigger supernova explosions, as some research suggests \cite{janka2012explosion}. Otherwise, this factor must be included in our analysis.

It should also be noted that both of the BBN and supernova considerations are updates of the original anthropic argument of \cite{carr1979anthropic}, which instead arrived at the condition $(m_e/M_{pl})^{1/2}\sim m_e^2/\vEW^2$, or $\rho_\text{EW}\sim\beta^{3/4}\gamma^{3/4}$. This additional factor of $\beta$ does not strongly affect our considerations, as its anthropic range is much smaller than that of $\gamma$.

Finally, we note that in \cite{harnik2006universe} it was proposed that a universe with very large $\vEW$ may still be habitable (even with $\rho_\text{EW}\sim1$), as long as the Yukawa constants are suitably adjusted to maintain the quark and lepton mass values, and cosmological parameters are adjusted to prevent all hydrogen from being depleted during BBN. However, in universes such as these an alternative to stellar nucleosynthesis must be found in order for significant oxygen to be present \cite{clavelli2006problems}.

\subsection{GUTs}

Ordinarily, the weak coupling has no bearing on the macroscopic world (since weak decays depend only on combinations of $\alpha_w$ and the W or Z boson masses so as to cancel the $\alpha_w$ dependence in favor of $\vEW$ \cite{peskin2018introduction}), and so the integral $\int d\alpha_w$ only contributes an irrelevant prefactor to the measure. One exception to this is in the instance of grand unified theories (GUTs), that impose the condition that the strengths of the electric, weak and strong forces (which depend on the energy scale used to measure them in quantum field theory) are equal at some high energy scale, where unification takes place (see for example \cite{zee2010quantum}). This enforces a constraint on the low energy couplings,
\beq
\frac{1}{\alpha_\text{QCD}}+\frac{c_\text{EM}}{\alpha}-\frac{c_w}{\alpha_w}=\Delta_3\label{gut}
\eeq
The constants $c_\text{EM}$ and $c_w$ are $\mathcal O(1)$ coefficients that depend on the particle content of the specific GUT theory (for instance, in the supersymmetric SU(5) theory $c_\text{EM}=3/7$ and $c_w=15/7$ \cite{particle2022review}. The factor $\Delta_3$ is equal to 0 to leading order, but with subleading corrections it is generically related to the logarithmic ratio of the masses of two heavy particles, $\Delta_3=c_3\log(M_{X_1}/M_{X_2})$. Again in supersymmetric SU(5), $c_3=19/(14\pi)$.

Equation \ref{gut} is implied to be evaluated at some energy scale, which is conventionally taken to be the Z boson mass. However, to eliminate the strong coupling $\alpha_\text{QCD}$ in favor of the more physically relevant variable $\gamma$, we may take this relation to apply at the proton mass. From the discussion on dimensional transmutation in the beginning of this section, we then have $1/\alpha_\text{QCD}=-9/(2\pi)\log(\gamma)+\text{const}$, where the constant may be absorbed into the definition of $\Delta_3$. 

We now describe two variants of GUT theory, which have dramatically different consequences for the resulting induced weight function. In the first scenario, the coefficient $\Delta_3$ is fixed by fundamental theory, and so has no freedom to vary. In this case the contribution to the measure from the weak sector is
\beq
W_\text{weak} = \int_0^1d\alpha_w\,\delta\left(-\frac{9}{2\pi}\log\gamma+\frac{c_\text{EM}}{\alpha}-\frac{c_w}{\alpha_w}-\Delta_3\right)
\eeq
where $\delta(x)$ is the Dirac delta function, enforcing the GUT condition. Because the integrand does not explicitly depend on $\alpha_w$, this evaluates to 1 (for all relevant values of $\alpha$ and $\gamma$. In this instance, the fact that the fundamental theory is a GUT does not directly influence the multiverse measure.

In the second scenario the details of the high energy physics are not fixed by any fundamental principle, and so the factor $\Delta_3$ can vary. For definiteness, if we take the masses of the two particles involved in its definition above to be uniformly distributed, $m_{X_i}\sim\mathcal U(0,\LUV)$, we find that the distribution of $\Delta_3$ is $p(\Delta_3)=1/(2c_3) e^{-|\Delta_3|/c_3}$. In this case, the weak contribution to the measure is
\bea
W_\text{weak} &=& \int_0^1d\alpha_w\,\int_\mathbb R d\Delta_3 p(\Delta_3)\,\delta\left(-\frac{9}{2\pi}\log\gamma+\frac{c_\text{EM}}{\alpha}-\frac{c_w}{\alpha_w}-\Delta_3\right)\nonumber\\
&=&\int_0^1d\alpha_w\,\frac{1}{2c_3}\,\exp\left(\frac{-1}{c_3}\left|-\frac{9}{2\pi}\log\gamma+\frac{c_\text{EM}}{\alpha}-\frac{c_w}{\alpha_w}\right|\right)\nonumber\\
&\propto&\gamma^{\frac{9}{2\pi c_3}}\,\exp\left(-\frac{c_\text{EM}}{c_3\,\alpha}\right)
\eea
In this case, the GUT constraint induces a nontrivial extra factor in the measure, involving both $\alpha$ and $\gamma$.

\subsection{Higgs-top sector}

The masses of the Higgs boson and top quark have no overt bearing on any of the macroscopic properties of the world. However, one very important effect they have a strong influence on is the stability of the standard model vacuum, which is in fact not guaranteed. In \cite{isidori2001metastability} it was found that our vacuum is only stable if the following condition holds:
\beq
\sigma_H>k_t\Delta_t-k_s\alpha_\text{QCD}-k_0\label{higgsstab}
\eeq
Where $k_t=2$, $k_s=1600\pi  s/9$ and $k_0=243.0 s$, with $ s = \text{GeV}/M_{pl}=4.11\times10^{-19}$. If this condition had not held, the Higgs field would have been susceptible to spontaneously tunneling away from its current vacuum expectation value $\vEW$ to very high values, destabilizing all other particles along with it in rapidly expanding bubbles that would destroy our known universe. Alternatively, we note that this stability criterion is sensitive to any heavier unknown particles that couple to the Higgs, and so it may be that stability is guaranteed in the completion of the standard model \cite{hiller2024vacuum}.

In \cite{froggatt1996standard} the principle of criticality was introduced, which states that the universe should be not quite stable, but instead \emph{metastable}- that is, it should exist on the threshold of stability so that the typical time it takes for bubbles to nucleate should be exceedingly long. Interestingly, this principle correctly predicted both the Higgs and top masses before they were discovered. Therefore, we may rightly consider both scenarios in a multiverse context, in which equation \ref{higgsstab} holds as an approximate equality (metastability) or where it holds as an equality (stability) \footnote{In the metastable case, the coefficients are instead $k_t=2.9$, $k_s=2500\pi s/9$, and $k_0=122.6 s$}.

To compute the induced weight in these scenarios, we may replace the strong force coupling with the ratio of the proton mass to Planck mass through $\alpha_\text{QCD}=-2\pi/(9\log(c\gamma))$ in equation \ref{higgsstab}, as before. Then, in the metastability case, this region of phase space can be integrated as
\bea
W_\text{Ht} &=& \int_0^{q_{pl}^{-1}}\frac{d\Delta_t}{\Delta_t}\int_0^{q_{pl}^{-1}}d\sigma_H\,\sigma_H\,\delta\left(\sigma_H-k_t\Delta_t-\frac{k_s}{\log \tilde c\gamma}+k_0\right)\nonumber\\
&=& \text{const}+\frac{B}{\log \tilde c \gamma}
\eea
This only depends logarithmically on the constant $\gamma$, and so we neglect this contribution, as before.

The case where true stability is enforced is more cumbersome but with similar conclusions. The induced weight then becomes
\bea
W_\text{Ht} &=& \int_0^{q_{pl}^{-1}}\frac{d\Delta_t}{\Delta_t}\int_0^{q_{pl}^{-1}}d\sigma_H\,\sigma_H\,\theta\left(\sigma_H-k_t\Delta_t-\frac{k_s}{\log \tilde c\gamma}+k_0\right)\nonumber\\
&=& \int_0^{q_{pl}^{-1}}d\sigma_H\,\sigma_H\,\frac{1}{\log q_\text{IR}/q_\text{UV}}\log\left(\frac{\sigma_H-\frac{k_s}{\log \tilde c\gamma}- k_0}{k_t\,q_\text{IR}^{-1}}\right)
\eea
We will spare the reader the full expression for this integral, but again this depends only logarithmically on $\gamma$. Therefore, we can conclude that vacuum stability considerations do not exert enough constraint on parameter space to substantially alter any of the expectations we have for the macroscopic world.

The effects of including these various induced weights will be discussed in section \ref{results}. We summarize the forms of these induced weights in Table \ref{micro_table}.

\begin{table*}
    \centering
    \begin{tabular}{|c|c|}
        \hline
        condition & $W$ \\
        \hline
        H from BBN & const + $\mathcal O(\log)$ \\
        neutrino SN explosions & $\gamma^{-3/4}$ \\
        \hline
        rigid GUT & const  \\
        flexible GUT & $\gamma^{\frac{9}{2\pi c_3}}\,\exp\left(-\frac{c_\text{EM}}{c_3\,\alpha}\right)$\\
        \hline
        vacuum stability & const + $\mathcal O(\log)$ \\
        vacuum metastability & const + $\mathcal O(\log)$ \\
        \hline
    \end{tabular}
    \caption{Contribution to the induced weight for various microscopic conditions described in the text.}
    \label{micro_table}
\end{table*}

\section{Cosmological Variables}\label{cosmovar}

The standard model of cosmology is specified by even fewer variables than the standard model of particle physics. These are: the amplitude of primordial fluctuations $Q$, the baryon to photon ratio $\eta$, the dark matter density $\xi_\text{dm}$, and the vacuum energy density $\rho_\Lambda$ \cite{matters}. There are a few others, such as the spectral tilt of primordial fluctuations, primordial gravitational wave power, and spatial curvature which are not as influential and so we will not consider explicitly. Many of these quantities are very clearly composite, rather than fundamental constants. The cosmological measure for these variables can be written as
\beq
d\xcosmo\,  p_\text{cosmo}=dQ\,p_Q(Q)\,d\eta\, p_\eta(\eta)\,d\xi_\text{dm}\, p_\text{dm}(\xi_\text{dm})\,d\rho_\Lambda\, p_\Lambda(\rho_\Lambda)\,d\mu_\text{irrel}
\eeq

These variables are closest what appear in the underlying theory, and so it will be easiest to determine priors for these. However, they are not necessarily the most convenient when determining anthropic bounds and how they influence habitability, so throughout we will transform to more convenient final variables. For instance, the dark matter to baryon ratio $\omega$ will ultimately be favored over $\xi_\text{dm}$. The composite variable $\kappa=Q(\omega\eta)^{4/3}$, which dictates the density of galaxies, will function as a macroscopic variable, as it has influence over a number of different aspects of galactic habitability.

In equation \ref{Wcosmo} above, the induced weight is dependent on three factors: The cosmological measure $ p_\text{cosmo}$, which depends on assumptions made about the physics of various cosmological scenarios, the number of protons per universe $N_{\frac{\text{protons}}{\text{universe}}}$, and the star formation efficiency $\epsilon_\text{SF}$. These are then incorporated into the cosmological induced weight. After this, we discuss several contributions to galactic habitability.


\subsection{Cosmological Priors}

Since the physics that dictates the cosmological parameters is still unknown, most of their prior probability distributions depend sensitively on the theories that have been proposed to explain them. 
Below we outline some of the most popular theories for these variables, and the resulting prior probability distributions.\\

\noindent{\bf Cosmological Constant\\}
The prior for the vacuum energy density is actually fairly insensitive to the assumptions about the ultimate physics which dictates this quantity. As discussed in \cite{cc}, the prior for the cosmological constant $\rho_\Lambda$ is expected to be flat in the neighborhood of 0, because this value is not special from a microscopic perspective. So, we can treat the cosmological constant as governed by $\rho_\Lambda\sim\mU(-\LUV^4,\LUV^4)$, particularly in the comparatively narrow range of anthropically allowed values. The fact that the observed value is so extraordinarily small compared to typical values of this distribution is the usual cosmological constant problem. We may define a dimensionless quantity $\hat\rho_\Lambda=\rho_\Lambda/M_{pl}^4$; in terms of this variable, Weinberg in the above reference found the anthropic bound $\hat\rho_\Lambda<\kappa^3\gamma^4$ in order for galaxies to form.\\

\noindent{\bf Dark Matter\\}
There are a number of different theories as to the nature of dark matter, and each of these has a different prior for the dark matter abundance. The dark matter abundance is usefully parameterized as $\xi_\text{dm}=\rho_\text{dm}/(n_\gamma M_{pl})$, which is a dimensionless ratio of the dark matter abundance to photon density- a quantity that is constant in time through cosmic evolution \cite{Kolb:1990vq}. We will ultimately want to express this through the alternate variable $\omega=\rho_\text{matter}/\rho_\text{baryon}=1+\xi_\text{dm}/(\eta\gamma)$, as this variable disentangles the dark matter sector from the macroscopic variables. In \cite{matters} it was stated that if $\omega\lesssim3.5$, severe Silk damping would prevent galaxies from forming. This reference contains other bounds on the dark matter abundance, but these are all less important and more controversial \cite{bousso2013comparable}.

We now detail the priors for a number of leading dark matter candidates.
\begin{itemize}
\item \textbf{Freeze-out}: In the freeze-out scenario, dark matter is initially in equilibrium, but as the universe expands, a threshold is crossed where dark matter annihilation processes cease to occur, leaving a relic abundance. In this scenario, the dark matter abundance is given by $\xi_\text{dm}\sim m_\text{dm}^2/(\alpha_\text{dm}^2M_{pl}^2)$ \cite{feng2023wimp}. If we assume the prior distributions for the dark matter mass squared is uniform, $m_\text{dm}^2\sim\mathcal U(0,\Lambda_\text{UV}^2)$, we can rewrite the theory prior as $dm_\text{dm}^2/M_{pl}^2\,d\alpha_\text{dm}=\alpha_\text{dm}^2d\alpha_\text{dm}d\xi_\text{dm}$. In the case where the dark matter force is separate from all other physics, the coupling strength can be integrated out to no effect, and we are left with a uniform prior for $\xi_\text{dm}$.
\item \textbf{WIMP}: This is the same scenario as freeze-out above, except that in this case the dark matter force is the same as the weak force of the standard model (at least, as it is defined here). This scenario is motivated by the \emph{WIMP miracle}, in that it quite naturally results in having the observed dark matter abundance \cite{Kolb:1990vq}. We make a distinction from more general freeze-out models here because WIMPs may result in a different induced weight. If we do not assume the GUT paradigm, the force strength can again be integrated over to no effect, and the resulting prior is the same as in the freeze-out scenario above. If we do assume the GUT paradigm, then we must take the constraint from equation \ref{gut} into account. In this case our split between cosmological and microscopic variables breaks down, and the induced weight picks up an additional factor. For the case of a rigid GUT, this is
\beq
W_\text{GUT+WIMP}=\left(\frac{c_w}{-\frac{9}{2\pi}\log\gamma+\frac{c_\text{EM}}{\alpha}-\Delta_3}\right)^2\label{gutwimp}
\eeq
The prior for dark matter mass is still uniform in this case. We do not endeavor to calculate the induced weight for the case of WIMPs and flexible GUTs here.
\item \textbf{Axion}: Axions are a well-motivated dark matter candidate because they are expected to be present from a theoretical perspective and are the simplest solution to the \emph{strong CP problem}, which is the absence of certain parity-violating strong force processes \cite{svrcek2006axions}. In this scenario, dark matter is a condensate of the axion field, rather than an aggregate of particles. The axion dark matter abundance is given by $\rho_\text{dm}\sim f_a^4 \sin^2\theta_0$, where $f_a$ is the axion coupling constant and $\theta_0$ is an initial misalignment angle \cite{marsh2016axion}. The theory prior is log-uniform for $f_a$, and uniform for $\theta_0$ \cite{arvanitaki2010string}. There are two possible scenarios for axion dark matter- in the first, superhorizon scenario, the scale of $f_a$ is very high, leading to a dark matter abundance that is different for different patches throughout our larger universe and naturally much larger than our observed abundance. In this scenario, the initial misalignment angle in our patch is concomitantly small to satisfy the anthropic bound on galaxy formation, leading to the observed abundance. In this case, $\sin\theta_0\sim\theta_0$, $\theta_0\sim1/2\sqrt{n_\gamma/(\xi_\text{dm}f_a^4)}$, and $p_\text{dm}(\xi_\text{dm})\sim1/\sqrt{\xi_\text{dm}}$, as derived in \cite{matters}. This holds true whether $f_a$ is held fixed or is allowed to scan in the fundamental theory.

In the alternate subhorizon scenario, $\theta_0$ varies on small scales, and the dark matter abundance is given by an average over these regions, so that $\sin^2\theta_0\sim1/2$. Here $f_a\sim(2\xi_\text{dm}n_\gamma)^{1/4}$ and $p_\text{dm}(\xi_\text{dm})\sim1/\xi_\text{dm}$.

\item \textbf{Primordial Black Holes}: Lastly, we discuss primordial black holes, which result whenever processes in the early universe are violent enough to cause regions of spacetime to collapse \cite{green2021primordial}. In this scenario, regions which are above some density threshold collapse- because this is set by a Gaussian process, the dark matter abundance depends exponentially on some of the parameters in the model. This results in a log-uniform distribution for the abundance, $p_\text{dm}(\xi_\text{dm})\sim1/\xi_\text{dm}$. 

In passing, we also mention that freeze-in dark matter, where production is always far from equilibrium, can also produce a log-uniform abundance distribution. This is the case for gravitationally produced dark matter, as in \cite{garny2016planckian}, again because the theory depends exponentially on some parameters. In the case where freeze-in dark matter is produced by another `feeble' force, we have $\xi_\text{dm}\sim\lambda_f^2$, where $\lambda_f$ is a coupling in the theory \cite{hall2010freeze}. When this is a Yukawa coupling, the resulting distribution will be log-uniform as well.
\end{itemize}

In summary, we may say that $p_\text{dm}(\xi_\text{dm})\sim\xi_\text{dm}^{q_\text{dm}}$, where $q_\text{dm}$=0 for the freeze-out and WIMP scenarios (which we denote {\bf FO}), -1/2 for superhorizon axions (denoted {\bf supA}), and -1 for subhorizon axions, primordial black holes, and freeze-in scenarios (denoted {\bf SI} for Scale Invariant). Then the dark matter prior can be written as
\beq
d\xi_\text{dm}\,p_\text{dm}(\xi_\text{dm})=d\omega\,(\eta\gamma)^{1+q_\text{dm}}(\omega-1)^{q_\text{dm}}\label{pdmxfm}
\eeq

\noindent{\bf Baryon to Photon Ratio\\}
The baryon to photon ratio is the result of some fundamental preference for matter over antimatter which resulted in a very slight overabundance of matter in our universe. The ultimate source of this asymmetry in the early universe is unknown and there is little hope of probing this mechanism further in the immediate future. Consequently there are a great many different theories for baryogenesis; here, we cover a few of the more popular ones, which serve as a representative sample for the range of possibilities.
\begin{itemize}
\item \textbf{Leptogenesis}: Is a theory of the origin of baryon asymmetry that is mediated through an initial phase of lepton number violation \cite{davidson2008leptogenesis}. These scenarios are motivated by the need for new physics in the neutrino sector to account for their masses, and employ an additional heavy particle that induces lepton and baryon number violation. In these scenarios, the resultant baryon to photon ratio is proportional to the square of a Yukawa coupling in the theory, $\eta\sim \lambda_L^2$ \cite{barbieri2000baryogenesis}. If we take the Yukawa coupling prior distribution to be log-uniform, then the resulting distribution for the baryon to photon ratio will also be log-uniform, $p_\eta(\eta)\sim1/\eta$.
\item \textbf{Affleck-Dine}: In this scenario, the baryon asymmetry is induced by the vacuum expectation value of some heavy field, usually considered to be the superpartner of a combination of standard model particles \cite{affleck1985new}. In this scenario, the baryon asymmetry is related to the initial field value as $\eta\sim \phi_0^2/\LUV^2$. If the field value $\phi_0$ is uniformly distributed, then the resulting distribution for $\eta$ is $p_\eta(\eta)\sim 1/\eta^{1/2}$. Though in this scenario the resulting baryon asymmetry is typically much larger than our observed value, anthropic effects could be responsible for enforcing our small observed value \cite{linde1985new}.
\item \textbf{Electroweak baryogenesis}: Though the standard model is incapable of producing a baryon asymmetry at the levels observed, it is possible to obtain the observed baryon to photon ratio in some extensions of the standard model \cite{morrissey2012electroweak}. As these models rely on nonperturbative processes to generate the baryon asymmetry, the resulting value will depend exponentially on model parameters \cite{trodden1999electroweak}. As such, the distribution of $\eta$ will be approximately log-uniform, $p(\eta)\sim \eta^{-1}$.
\end{itemize}

The anthropic bounds on $\eta$ directly are quite weak- \cite{nanopoulos1980bounds} mentions that $\eta$ may be constrained to be within $10^{-11}-10^{-4}$, compared to the observed value of $10^{-9}$. Many bounds on this parameter are actually induced through bounds on the galactic density, $\kappa$. As such, we favor eliminating this variable in favor of $\kappa$ via
\beq
d\eta\,p_\eta(\eta) = \frac{d\kappa}{Q^{3/4}\omega\kappa^{1/4}}\,p_\eta\left(\frac{\kappa^{3/4}}{Q^{3/4}\omega}\right)\label{petaxfm}
\eeq
In this paper bounds on this quantity are enforced as they manifest on the galactic density, discussed later in this section.

In summary, we have $p_\eta(\eta)\sim\eta^{q_\eta}$, where $q_\eta=-1$ for low energy baryogenesis (the leptogenesis and electroweak baryogenesis scenarios, denoted {\bf lowE}) and $q_\eta=-1/2$ for high energy baryogenesis (the Affleck-Dine scenario, denoted {\bf highE}).\\
    
\noindent{\bf Density Perturbations\\}
The amplitude of primordial density perturbations $Q$ is thought to be set by quantum fluctuations of some field in the early universe, most commonly associated with the inflaton \cite{starobinsky1982dynamics}. The observed value is about $10^{-5}$, and the anthropic range for this quantity was found to be $10^{-6}\lesssim Q\lesssim 10^{-4}$ \cite{starstar}. As such, observations favor a scenario where the distribution of values is approximately log-uniform, including the prior and any additional contributions to the weight from other sectors. However, in many models this is not the case, where instead either small or large values of $Q$ are strongly preferred- a problem known as the \emph{Q catastrophe} \cite{hep-th/0508005}, where it was shown to also depend strongly on cosmological measure taken. We do not outline the priors in any explicit theories, because for the most part this term remains isolated from the rest of the parameters we discuss, but we display the resulting posterior distribution below, and note that this issue should be useful in determining which theories of the early universe are worthwhile.

\subsection{Number of Protons per Universe}
We now concern ourselves with the factor $N_{\frac{\text{protons}}{\text{universe}}}$. This quantity is notoriously tricky to handle, as it is formally infinite in an open or flat universe. In order to compare between different universes, then, a regularization technique is needed, but the results depend sensitively, in some cases to an extreme degree, on the choice of regularization used- this is the \emph{measure problem} \cite{Fmeasure}. Part of the subtlety lies in the fact that each universe will have a unique thermal history specified by its particle content, so finding a common reference point to fairly compare two different universes is difficult. In this paper we make use of the scale factor cutoff measure, which is quite naturally motivated, is compatible with our observed value of the cosmological constant, and avoids the more blatant problems that afflict earlier proposals of measure \cite{scale}.

We may define the volume of a patch as $V=4\pi/3 (a(t)\chi(t))^3$, with cosmological scale factor $a(t)$ and comoving distance 
\beq
\chi(t)=\int_0^t\frac{dt}{a(t)}
\eeq

The time evolution of the scale factor is dictated by the Friedmann equation
\beq
H^2=\frac{\dot a^2}{a^2}=\frac{8\pi}{3}G\left(\frac{\rho_\text{radiation}}{a^4}+\frac{\rho_\text{matter}}{a^3}+\rho_\Lambda\right)
\eeq
After matter-radiation equality, which in our universe occurred around 50 kyr, the first term can be neglected. With this approximation this equation can be solved analytically to yield
\beq
a(t)=\left(\frac{H_\text{matter}}{H_\Lambda}\right)^{2/3}\sinh\left(\frac32H_\Lambda t\right)^{2/3}
\eeq
Where we have defined $H_i=\sqrt{8\pi G\rho_i/3}$.

Then the comoving distance can be written in terms of the hypergeometric function as 

\beq
\chi(t)=\frac{2}{H_m^{2/3}H_\Lambda^{1/3}}\sinh\left(\frac32H_\Lambda t\right)^{1/3}{}_2F_1\left(\frac16,\frac12;\frac76;-\sinh\left(\frac32H_\Lambda t\right)^2\right)
\eeq
At early and late times, this asymptotes to
\beq
\chi(t)\rightarrow
\begin{cases}
    (12t)^{1/3}H_m^{-2/3},& t\ll H_\Lambda^{-1}\\
    2\Gamma(1/3)\Gamma(7/6)/(\sqrt{\pi}H_m^{2/3}H_\Lambda^{1/3}), & t\gg H_\Lambda^{-1}
\end{cases}
\eeq

Since the proton density is given by $\rho_m/(\omega a(t)^3)$, the number of protons per patch is then
\beq
N_{\frac{\text{protons}}{\text{universe}}}(t) = \frac{4\pi}{3}\frac{\rho_m}{\omega}\chi(t)^3\sim48\pi\frac{M_{pl}^2}{\omega\,m_p}\min\left(t,\frac{1.8}{H_\Lambda}\right)\label{Nprotu}
\eeq

This quantity depends on time, but for the total number of protons per universe we use the asymptotic value. As a side note, this $1/\omega$ dependence was found in \cite{freivogel2010anthropic}, where it was shown to be important for suppressing any natural preference for a higher abundance of dark matter. This accounts for the observed fact that the observed abundance ratio is close to the anthropic lower bound.

\subsection{Cosmological Induced Weight}
With these ingredients, we can arrive at the expression for the cosmological induced weight. From equation \ref{Wcosmo}, this is defined as
\beq
W_\text{cosmo}(\xmacro)=\int dQ\,d\eta\,d\xi_\text{dm}\,d\rho_\Lambda\,p_Q(Q)\,p_\eta(\eta)\,p_\text{dm}(\xi_\text{dm})\,
N_{\frac{\text{protons}}{\text{universe}}}\,
\epsilon_\text{SF}
\eeq




Then using \ref{pdmxfm} for $p_\text{dm}(\xi_\text{dm})$, \ref{petaxfm} for $p_\eta(\eta)$, and \ref{Nprotu} for $N_{\frac{\text{protons}}{\text{universe}}}$, we have

\beq
W_\text{cosmo}=
\frac34\, \int dQ\,\frac{\left(\frac{\kappa^{3/4}}{Q^{3/4}\,\omega}\right)^{q_\eta}\,
p_Q(Q)}{Q^{3/4}\,\omega\,\kappa^{1/4}}\,
\int_{3.5}^\infty d\omega\, \left(\frac{\kappa^{3/4}\,\gamma}{Q^{3/4}\,\omega}\right)^{1+\qdm}(\omega-1)^{\qdm}\,
\int_0^{\kappa^3\gamma^4}d\hat\rho_\Lambda\,
\frac{48\pi1.8}{\omega\,\gamma\,\hat\rho_\Lambda^{1/2}}\,
\epsilon_\text{SF}
\eeq

Note the posterior for $\hat\rho_\Lambda\propto\hat\rho_\Lambda^{-1/2}$ as a consequence of the number of protons in a horizon volume being proportional to this factor. This additional induced dependence differs from the usual flat prior that is found in the literature, and favors universes with smaller cosmological constant as those have more protons. Usual treatments take the probability of being in a particular universe to be proportional to the fraction of baryons that end up in sufficiently large galaxies, and so do not include this factor, following \cite{cc}. In \cite{sorini2024impact} it was found that including a factor akin to this can alleviate some tensions in  the probability of observing our value of $\rho_\Lambda$\footnote{Note that if we instead treat $N_{\frac{\text{protons}}{\text{universe}}}$ as a function of time, for $t<1/H_\Lambda$ this dependence is not induced. If we then evaluate at the characteristic galaxy free-fall timescale $t_\text{ff}\sim1/\sqrt{G\rho}$, the resulting dependence of $W_\text{cosmo}$ on the macroscopic parameters is identical to the expression above. This is a consequence of the condition $t_\text{ff}<1/H_\Lambda$ being equivalent in form to the Weinberg bound on $\rho_\Lambda$.}.

The probability of observing our value of $\rho_\Lambda$ has been the subject of much discussion \cite{GLV}, $\,\,\,\,\,\,\,\,$ \cite{hep-th/0508005}, $\,\,\,\,\,\,\,\,$ \cite{astro-ph/0611573}, \cite{scale}, \cite{1508.01034}, \cite{sorini2024impact}. This quantity influences star formation through its influence on galaxy formation and evolution, but it has no effect on planetary habitability. Given the historical (and ongoing) treatment of this variable and its relative inactivity for determining habitability, we do not focus on it here, but instead integrate it out with the approximation that $\epsilon_\text{SF}$ is independent of $\rho_\Lambda$. This then acts as a contribution to the induced weight for the other macroscopic variables, which can then be factored as
\beq
W_\text{cosmo}=
129.6\pi\,\kappa^{2+\frac34(q_\eta+\qdm)}\,
\gamma^{2+\qdm}\,
\left[\int\frac{dQ\,p_Q(Q)}{Q^{\frac32+\frac34(q_\eta+\qdm)}}\right]\,
\left[\int_{3.5}^\infty d\omega\, \frac{(\omega-1)^{\qdm}}{\omega^{3+q_\eta+\qdm}}\right]\,
\epsilon_\text{SF}
\eeq


The dependence on both $Q$ and $\omega$ factor into isolated integrals given in brackets above, again with the approximation that $\epsilon_\text{SF}$ does not depend on either of these quantities. For the induced weight, these can be treated as irrelevant prefactors. However, one may also compute the probability of observing our particular values. For $Q$, this depends on the function $p_Q(Q)$, which depends sensitively on the theory generating these perturbations. We only note that we expect the ultimate dependence on $Q$ to be approximately log-uniform, given that our observed value is logarithmically about midway through the anthropically observed range. For the dark matter to baryon ratio, we can explicitly compute the quantity $P_\text{dm}=P(\omega_\text{obs}\leq6)$ for our various theories. The exact value will depend on both $q_\eta$ and $\qdm$, as outlined in Table \ref{cosmo_table} below, but for all potential values we discussed above, the probability lies within .4-.6.

The last thing to note is that this induces a dependence on the strength of gravity, $W_\text{cosmo}\propto\gamma^{2+\qdm}$. This factor was not taken into account in our previous works, and so alters the quantitative and sometimes qualitative conclusions we had arrived at previously. However, many of the theories for dark matter have $\qdm=-1$, which if taken in conjunction with the SN neutrino condition in equation \ref{WEWSN} leads to an overall $\gamma$ dependence $W\propto \gamma^{1/4}$, which hardly alters our quantitative results.

The dependence of $W_\text{cosmo}$ on $\gamma$ and $\kappa$ for the various underlying theories we consider are displayed in Table \ref{cosmo_table}.

\begin{table*}
    \centering
    \begin{tabular}{|c|c|c|c|c|}
        \hline
        scenario & & & & $W\sim \kappa^{2+\frac34(q_\eta+\qdm)}\,\gamma^{2+\qdm+q_\text{sn}}$ \\
        \hline
        &$q_\text{dm}$ & $q_\eta$ & $P_\text{dm}$ & $q_\text{sn}=-3/4,\quad q_\text{sn}=0$ \\
        \hline

        SI + lowE & -1 & -1 & .46 & $\kappa^{1/2}\,\gamma^{1/4},\quad\kappa^{1/2}\,\gamma$\\
        SI + highE & -1 & -1/2 & .59 & $\kappa^{7/8}\,\gamma^{1/4},\quad\kappa^{7/8}\,\gamma$\\
        supA + lowE & -1/2 & -1 & .44 & $\kappa^{7/8}\,\gamma^{3/4},\quad\kappa^{7/8}\,\gamma^{3/2}$\\
        supA + highE & -1/2 & -1/2 & .57 & $\kappa^{5/4}\,\gamma^{3/4},\quad\kappa^{5/4}\,\gamma^{3/2}$\\        
        FO + lowE & 0 & -1 & .42 & $\kappa^{5/4}\,\gamma^{5/4},\quad\kappa^{5/4}\,\gamma^{2}$\\
        FO + highE & 0 & -1/2 & .56 & $\kappa^{13/8}\,\gamma^{5/4},\quad\kappa^{13/8}\,\gamma^{2}$\\
        \hline
    \end{tabular}
    \caption{Contribution to the induced weight for various microscopic conditions described in the text. Dark matter scenarios: FO (freeze-out), supA (superhorizon axion), SI (scale invariant: subhorizon axion, primordial black holes, or freeze-in). Baryogenesis scenarios: lowE (low energy: leptogenesis or electroweak baryogenesis), highE (Affleck-Dine baryogenesis). The parameter $q_\text{sn}$ dictates whether neutrino-driven supernova element production is important or not.}
    \label{cosmo_table}
\end{table*}

\subsection{Star formation efficiency}

The last factor we need to compute the cosmological induced weight is the total star formation efficiency $\epsilon_\text{SF}$. We will be most interested in the overall star formation efficiency- that is, the fraction of protons that ultimately end up in stars, rather than during a particular epoch of galactic evolution, and how this depends on constants.

Star formation history is a well studied subject in the multiverse context, beginning in earnest with \cite{bousso2010predictions}. This was improved upon in \cite{barnes2018galaxy}, which matched star formation histories to EAGLE simulations. In \cite{oh2022fate} the long term star formation history in the multiverse was considered, continuing far into the future. A more analytic treatment was undertaken in \cite{sorini2024impact}.

Star formation is a complicated process, and so treatments of star formation history in general tend to be semi-empirical, using galactic mass scales that are fit to observations and efficiency curves that fit the data well but are not necessarily well motivated from first principles. One approach we can take, since we only care about the ultimate fraction of protons that end up in stars, is to determine the star formation efficiency as a function of galaxy mass, and then integrate this against the galaxy mass distribution, $c(M)=\text{erfc}((M/M_\text{halo})^{2/3})$, where $M_\text{halo}=9.7*10^{-4}M_{pl}^3/(\eta^2\omega^2m_p^2)$, normalized to $9.7\times10^{11}M_\Sun$ \cite{ACBplanets}. 

In \cite{salcido2020feedback} total star formation as function of galaxy mass was described as an empirical double power law that peaks at $\sim10^{12}M_\Sun$. Generically, star formation can be thought of as nearly maximally efficient by default, with star formation suppressed in both galaxies which are too small or too large, for different reasons. In \cite{silk2012current}, it was argued that the cause of star formation suppression in small galaxies is supernova winds, and active galactic nuclei (AGN) jets in large galaxies. In \cite{behroozi2013average} it was found that approximately 2/3 of all star formation occurs within a factor of 3 of this peak halo mass.

Because the star formation efficiency curves take empirical forms, and the bulk of star formation occurs very close to the peak galactic mass, we can approximate the star formation efficiency as constant within this mass range and vanishing outside. This gives for the overall star formation efficiency:
\beq
\epsilon_\text{SF}\sim\epsilon_\text{SF}^0\left(\text{erfc}\left(\left(\frac{M_\text{lower}}{M_\text{halo}}\right)^{2/3}\right)-\text{erfc}\left(\left(\frac{M_\text{upper}}{M_\text{halo}}\right)^{2/3}\right)\right) 
\eeq
We now need to estimate the upper and lower bounds on galactic mass in this expression.

In \cite{dekel1986origin} it was found that the lower bound on halo mass is set by energy driven loss from supernovae. In this setup, the relevant quantity is the energy injection rate $\dot E_\text{SN}\sim \epsilon_\text{ISM}E_\text{SN}\Gamma_\text{SN}$, where $\Gamma_\text{SN}$ is the supernova rate and $\epsilon_\text{ISM}\sim.1$ is the efficiency of energy transfer to the interstellar medium. We can write the supernova rate as $\Gamma_\text{SN}\sim \epsilon_\text{SF}^0f_\text{SN}M_\text{galaxy}/M_\text{star}$, where $f_\text{II}$ is the fraction of stars that are large enough to become (type II) supernovae, which was argued in \cite{mc5} to be independent of constants. Star formation will become suppressed when the total energy injected rivals the galaxy's binding energy, $\dot E_\text{SN} t_\text{ff}\sim E_\text{bind}\sim G M_\text{galaxy}^{5/3}\rho^{1/3}$. Using $E_\text{SN}\sim\alpha M_\text{star}$, $t_\text{ff}\sim 1/\sqrt{G\rho}$ as the dynamical timescale set by the galaxy free fall time, and solving for $M_\text{galaxy}$, we have
\beq
M_\text{lower}=9.7\times10^{-9}\,\frac{\alpha^{3/2}\,M_{pl}^3}{\kappa^{3/2}\,m_p^2}
\eeq
Normalized to $7.9\times10^{10}M_\text{sun}$. When compared to the typical halo mass, we arrive at $M_\text{lower}/M_\text{halo}=9.9\times10^{-6} \alpha^{3/2}/Q^{3/2}$.\\


The upper limit on halo mass is instead given by AGN feedback. Following the King model for the central black hole mass of a galaxy, we have $M_\text{SMBH}=\omega \sigma_T \sigma_v^4/(2\pi G^2 m_p)$, where $\sigma_T$ is the Thomson cross section and $\sigma_v$ is the velocity dispersion of the galaxy \cite{king2003black}. What's of more relevance is the luminosity of the central black hole- as the Eddington luminosity sets the reference point, we have $L_\text{SMBH}=2\omega\sigma_v^4/G\sim G M_\text{halo}^{4/3}\rho_\text{halo}^{2/3}$, demonstrating that the mass of the central black hole scales superlinearly with halo mass, as indicated in \cite{haring2004black}. Lastly, the Salpeter timescale, which governs the black hole growth and is independent of mass, is given by $t_\text{Salpeter}\sim\sigma_T/(Gm_p)=1.86 \alpha^2M_{pl}^2/(m_e^2m_p)$, which has been normalized to $5\times10^7$ yr \cite{balbi2017habitability}.

The central AGN will play a significant role when it injects enough energy to the galaxy to unbind it- this is given when $L_\text{SMBH}t_\text{Salpeter}\sim M_\text{halo}\phi_\text{halo}$ \cite{buchner2024impediments}. The two are equal when $M\sim Q^3/(\omega^3 G^3 \rho_\text{halo}^2 t_\text{Salpter}^3)$, or
\beq
M_\text{upper}=7.5\times10^{-5}\,\frac{Q^3}{\omega^3}\,\frac{m_e^6}{\kappa^6\,\alpha^6\,m_p^5}
\eeq
Normalized to $4.0\times10^{13}M_\Sun$. In terms of a ratio to the average galaxy mass, we have $M_\text{upper}/M_\text{halo}=.078Q^{3/2}\beta^6\gamma^3/(\kappa^{9/2}\alpha^6)$.

Our final expression for the star formation efficiency is then 
\beq
\epsilon_\text{SF}=\epsilon_\text{SF}^0\left(\text{erfc}\left(4.6\times10^{-4}\,\frac{\alpha}{Q}\right)-\text{erfc}\left(.18\,\frac{Q\,\beta^4\,\gamma^2}{\kappa^3\,\alpha^4}\right)\right)\label{sffinal}
\eeq
With the understanding that this is equal to 0 when $M_\text{lower}>M_\text{upper}$, which occurs when $1.3\times10^{-4}\alpha^{15/2}\kappa^{9/2}/(Q^3\beta^6\gamma^3)>1$.

\subsection{Galactic Habitability}\label{galhab}

Since we consider variations in the galactic density $\kappa$ within our calculations for the first time in this paper, we also need to include how this parameter can impact habitability. Indeed it is important, as galactic density controls the rates of a number of phenomena that could adversely affect life. Here, we focus on three: stellar encounters, supernova explosions, and active galactic nuclei. Additionally, galactic density influences a number of planetary properties indirectly, chiefly through its impact on the initial size of protoplanetary disks \cite{mc2}. Here we derive expressions for the various components of galactic habitability, to incorporate into our analysis in the next section. These will all formally be a part of the habitability $\mathbb H$ in equation \ref{eqnOWW}.\\

\noindent{\bf Stellar encounters\\}
Passing stars could severely disrupt planetary systems, precluding them from being habitable. The denser a galaxy is, the greater number of encounters, and so this effect diminishes the habitability of regions of parameter space where galaxies are very dense. The fraction of unaffected systems can be written as $p_\text{survive}=e^{-\tau}$, with $\tau\sim n \sigma v t$, where $n$ is stellar density, $\sigma\sim \pi a_\text{temp}^2$ is the cross section of a planetary system \cite{li2015cross}, $a_\text{temp}\sim \text{AU}$ is a typical temperate orbit \cite{mc2}, $v\sim\sqrt{E_\text{Rydberg}/m_p}$ is the typical galactic velocity dispersion \cite{mc2}, and $t\sim t_\star$ is the lifetime of the planetary system\footnote{Here we consider encounters throughout the lifetime of the planetary system, but encounters may in fact be much more common in stellar birth clusters due to their higher density, despite the fact that they're much shorter lived \cite{li2015cross}. In this scenario, density and cross section scale the same as above, but lifetime is set by the free-fall time of the cluster. Additionally, we do not consider the effects of gravitational enhancement on the cross section here, which would be relevant for slow-moving encounters.}. The optical depth $\tau$ is then
\beq
\tau \sim\frac{\kappa^3\,\lambda}{\alpha^{17/2}\,\beta^{19/4}}
\eeq
Here, $\lambda$ is a dimensionless quantity comparing a star's mass to the Chandrasekhar mass $(8\pi)^{3/2}M_{pl}^3/m_p^2$. The overall normalization of the optical depth is dependent on the details of the planetary system, what qualifies as a disruption, and can only be estimated through simulations. There is some disagreement in the literature on the actual importance of this process, so we treat the normalization as a free parameter that dictates $p_\text{survive}$. For instance, \cite{kaib2025influence} suggest that the fraction of planetary systems that suffer from stellar encounters is $.5\%$ over the span of the 5 billion years of our solar system's evolution, resulting in $p_\text{survive}\sim.995$. In this work, we vary the value of $p_\text{survive}$ for each of the three contributions to galactic habitability we consider to determine their relative importance on the resulting multiverse probabilities.\\

\noindent{\bf Supernova explosions\\}
Supernova explosions represent another potential limiting effect on habitability that depends on galactic density. These have already been considered in the multiverse context in \cite{totani2019lethal}. The overall effect of supernova explosions is uncertain, but they have been implicated by some to play a role in at least one of Earth's mass extinctions \cite{fields2020supernova}, and may have contributed to minor extinctions at the onset of the most recent set of ice ages \cite{thomas2016terrestrial}. The survival probability for planetary systems under threat of supernova bombardment can similarly be parameterized as $p_\text{survive}=e^{-\tau}$, but this time with $\tau\sim\Gamma_\text{SN}\,t_\star$. The supernova rate can be written as
\beq
\Gamma_\text{SN}\sim \frac{f_\text{II}\,\rho_\text{gal}\,r_\dagger^3}{t_\text{ff}\,M_\star}
\eeq
Where $f_\text{II}$ is the fraction of stars that become type II supernovae, and $r_\dagger$ is the \emph{supernova death radius}. The death radius can be estimated along the lines of \cite{mc4} as $r_\dagger\sim R_\text{terr} \sqrt{N_\text{SN}/N_\text{ozone}}$, where $N_\text{SN}$ is the number of photons emitted by a supernova and $N_\text{ozone}$ is the number of ozone molecules present in the planet's atmosphere that are required to be destroyed for the effects of the supernova to be felt. The number of photons produced by a supernova is set by the total number of protons, so that $N_\text{SN}\sim M_\star/m_p$, and the number of ozone particles is $N_\text{ozone}\sim \pi R_\text{terr}^2/\sigma$, $\sigma\sim a_\text{Bohr}^2$, yielding
\beq
\tau_\text{SN}\sim\frac{\kappa^{3/2}}{\lambda^2\,\alpha\,\beta^5\,\gamma^{5/2}}
\eeq
Again, we will vary the prefactor in this expression to explore how this influences multiverse probabilities for different values of $p_\text{survive}$.\\

\noindent{\bf Active Galactic Nuclei\\}
Galactic nuclei, when in their active phase, emit a significant amount of UV radiation that can sterilize nearby planetary surfaces and, in even more extreme scenarios, strip planets of their atmospheres \cite{gonzalez2005habitable}. Though most estimates place the AGN radius of influence to be a few parsecs, \cite{balbi2017habitability} concluded under some pessimistic assumptions that significant atmospheric loss can occur even up to a kiloparsec away. Still others have argued that this increased flux may be beneficial for the emergence of life, as it provides for fertile disequilibrium photochemistry \cite{lingam2019active}.

The condition for the AGN radius of influence can be found by demanding that the flux coming from the central AGN be below a certain threshold needed for the maintenance of biomolecules. This can be expressed as $L_\text{SMBH}/(4\pi r^2)<T_\text{mol}^3 E_\text{Rydberg}$. Solving for $r$ and using $L_\text{SMBH}\sim GM_\text{halo}^{4/3}\rho_\text{halo}^{2/3}$ to express this in terms of constants, we have
\beq
r_\text{AGN}=406\,\frac{Q\,m_p^{3/4}\,M_{pl}}{\alpha^4\,m_e^{11/4}}
\eeq
Which has been normalized to 43 pc. Though this only represents a small fraction of the galaxy, it is the densest fraction, and this quantity depends quite sensitively on some of the constants.

To determine the fraction of stars within $r_\text{AGN}$, we can use the S\'ersic profile for the density of stars, $\rho_\star(r)\propto e^{-r/r_\text{gal}}$. With this, and using the 2-dimensional thin disk approximation, the fraction of stars with orbits larger than $r_\text{AGN}$ is $p_\text{survive}=(1+r_\text{AGN}/r_\text{gal})e^{-r_\text{AGN}/r_\text{gal}}$.\\

There are additional ways galactic density may influence habitability other than the three we consider here. The Galactic Habitable Zone \cite{Gonzalez:2001hh}, narrowly conceived of as regarding metallicity, may restrict habitability to an annulus, or even preclude habitability for some values of the constants. Passage through giant molecular clouds may also impact habitability, and would also depend on the constants \cite{kokaia2019stellar}. Also, the relative abundance of spiral versus elliptical galaxies may have an impact of overall habitability of a universe if elliptical galaxies are much less habitable, though this complicates our analysis as this is time dependent \cite{whitmire2020habitability}.

\begin{figure}
    \centering
    \includegraphics[width=0.6\linewidth]{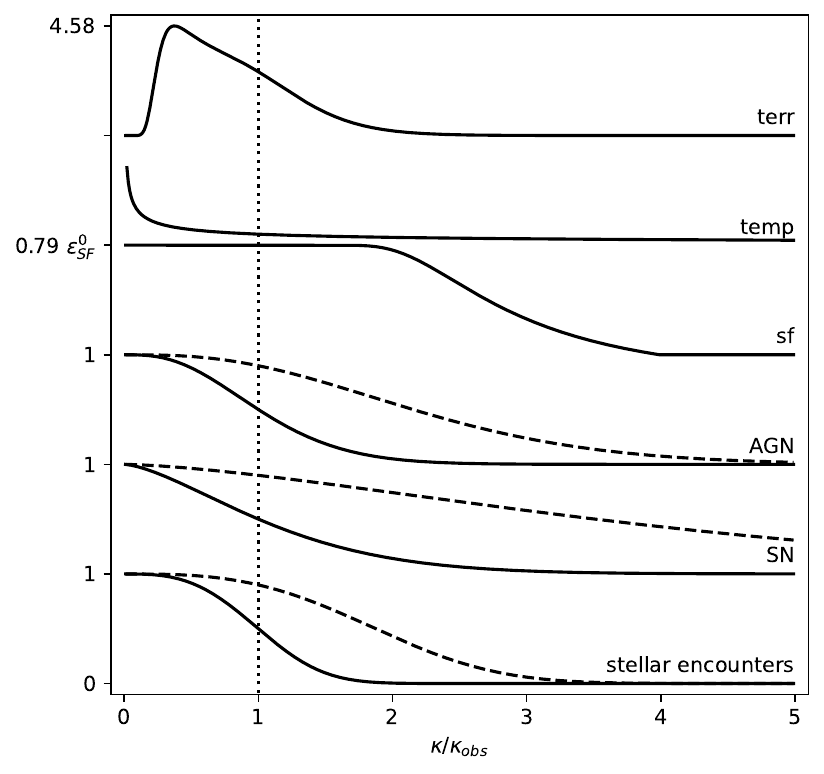}
    \caption{Dependence of various factors on the galactic density $\kappa$, with all other constants held fixed. The {\bf terr} and {\bf temp} curves are the average number of terrestrial and temperate planets per star, respectively. The {\bf sf} curve is the star formation efficiency compared to its maximum possible value. The last three curves denote the galactic habitability conditions discussed above. For these, solid curves correspond to $p_\text{survive}=.5$, and dashed to $p_\text{survive}=.9$. The {\bf temp} curve formally diverges as $\kappa\rightarrow0$, as it scales as $\kappa^{-1/2}$.}
    \label{kappadep}
\end{figure}

In Fig. \ref{kappadep} we display the dependence of these different types of disruptions on $\kappa$ for various choices of $p_\text{survive}$, along with some other relevant habitability conditions. In this figure, all other constants are held fixed at their observed values.

\section{Results}\label{results}

We can now marshal the expressions for fundamental physics and galactic habitability into our calculations of the probability of observing the various constants, $\alpha$, $\beta$, $\gamma$, $\delta_u$, $\delta_d$, and $\kappa$. We also include additional probabilities for non-fundamental quantities: the probability of orbiting a star at least as massive as our sun $P(\lambda)$, the probability of observing a Hoyle peak as large as ours $P(E_R)$, and the probability of observing such a large amount of organic material (carbon and oxygen), $P((C+O)/(Mg+Si))$ (see \cite{mc5}).

As shown in our previous works (cited below), these probabilities are highly sensitive to the assumptions we make on habitability conditions. These potential choices are numerous and span diverse considerations on environment and evolution, and so these conditions must be considered in combination. The full suite of combinations has already been unwieldy, and adding the 12 additional choices of weights from cosmological and fundamental physics only exacerbates this difficulty. Our primary aim is to determine which combinations of habitability conditions are compatible with the different assumptions made about fundamental and cosmological physics, and to determine trends among the different choices of combinations.

For all investigations, we include the star formation efficiency factor from equation \ref{sffinal}, which was previously assumed to be independent of the physical constants. For the optional habitability conditions, we take:
\begin{itemize}
\item {\bf photo}, {\bf yellow}: photosynthesis is required for habitability, with optimistic and pessimistic wavelength limits, respectively \cite{mc1}. 
\item {\bf TL}: planets are required to be tidally unlocked to be habitable \cite{mc1}.
\item {\bf bio}: only stars which burn for several billion years are considered habitable \cite{mc1}.
\item {\bf terr}: only terrestrial mass planets are habitable \cite{mc2}.
\item {\bf temp}: only planets orbiting in the the temperate zone are habitable \cite{mc2}.
\item {\bf time}: habitability is proportional to stellar lifetime \cite{mc3}.
\item {\bf area}: habitability is proportional to planet area \cite{mc3}.
\item {\bf S}: habitability is proportional to total disequilibrium produced, approximated as the total number of stellar photons that impinge on the planet's surface \cite{mc3}. 
\item {\bf C/O}, {\bf Mg/Si}: certain carbon-to-oxygen and magnesium-to-silicon ratios are required, respectively \cite{mc5}.
\item {\bf nitrogen}: appreciable nitrogen is required for habitability \cite{mc5}.
\item {\bf obliquity}: a large moon is required for habitability because it stabilizes planetary obliquity \cite{mc6}.
\item {\bf origin of life}: the probability of the origin of life on a planet is taken to be proportional to the total disequilibrium produced. Eleven scenarios for the origin of life are considered: lightning, solar energetic protons (SEP), extreme UV (XUV), hydrothermal vents (vents), interplanetary dust particles (IDP), comets, asteroids, a large impactor (moneta), interplanetary panspermia (plan pans), and interstellar panspermia (stel pans) \cite{mc8}.
\end{itemize}

Considering this full suite, especially in conjunction with the different theories of particle physics, cosmological scenarios, and galactic effects can be cumbersome. For exposition, we start simple and gradually add complexity to investigate the full set of effects.\\

\noindent{\bf Star formation significantly influences habitability\\}
Let's first determine the influence of including star formation on the probabilities we compute. To begin, we can use our previous minimal combination of habitability conditions that was compatible with observations, {\bf yellow S C/O}. Keeping galactic density fixed, the probabilities of observing our values of the macroscopic constants are
\beq
P(\alpha)=0.199, 
P(\beta)=0.293, 
P(\gamma)=0.122, 
P(\delta_u)=0.253, 
P(\delta_d)=0.324
\eeq

Including the effects of star formation, we have
\beq
P(\alpha)=0.206, 
P(\beta)=0.349, 
P(\gamma)=0.00971, 
P(\delta_u)=0.259,
P(\delta_d)=0.313
\eeq
So, while for most of the constants the effect is relatively minor, the probability of observing our value of the strength of gravity $\gamma$ is diminished by an order of magnitude when star formation is taken into account. This quantity is most sensitive because of its large allowed range.

However, we can find combinations of habitability conditions that are compatible with the multiverse when taking star formation into account. A minimal combination is to include the {\bf asteroid} condition as a source of disequilibrium preorganic material in origin of life scenarios. With this, the probabilities become 
\beq
P(\alpha)=0.364, 
P(\beta)=0.463,
P(\gamma)=0.189,
P(\delta_u)=0.357,
P(\delta_d)=0.341
\eeq
And so all probabilities are eminently compatible with our existence in this universe. This is interesting because the {\bf S} and {\bf asteroid} conditions in conjunction are incompatible with the multiverse when star formation is not taken into account \cite{mc8}.

This is not the only combination of conditions compatible with the multiverse. For instance, if the combination {\bf yellow TL temp Mg/Si moneta sf} is taken, the smallest probability is .227, making this combination even more preferred.\\

\noindent{\bf Galactic density constrains habitability conditions\\}
The above considered the effects of star formation only, to isolate the impact this has on the probabilities. We now include the effect of varying galactic density, which both induces additional parameter dependence, and also includes an additional probability, $P(\kappa)$, that can further constrain habitability conditions.

For now, we restrict our attention to the {\bf SI lowE SN} scenario of cosmology, where the distribution for both dark matter abundance and baryon to photon ratio are scale invariant and neutrinos are taken to be important for supernovae. This choice provides the weakest dependence on $\kappa$ and $\gamma$ of all the cosmology scenarios we consider. We will consider the alternative scenarios after this one. For this and all further results where galactic density varies, we also take the density perturbations $Q$ to vary log-uniformly, which as discussed in the previous section ensures that our observed value is typical.

The results for the probabilities discussed above are displayed in Table \ref{probs_table3} when including varying galactic density. For brevity, we've only included the variables which are impacted by the choices of habitability condition. The probabilities not displayed are all over .1. The same broad conclusions hold as in the case where galactic density is held fixed, with one notable exception- without taking star formation into account, the condition {\bf yellow S C/O} is untenable only because $P(\kappa)$ is small, demonstrating that considering this additional constant allows us to place more stringent constraints on habitability conditions than we otherwise could have obtained.\\



\begin{table*}
    \centering
    \begin{tabular}{|c|c|c|c|}
        \hline
        condition & $P(\beta)$ & $P(\gamma)$ & $P(\kappa)$\\
        \hline
        yellow + S + C/O & 0.293 & 0.122 & 0.0311\\
        yellow + S + C/O + sf & 0.434 & 0.00431 & 0.0417\\
        yellow + S + C/O + asteroids & 0.00161 & 0.00106 & 0.00126\\
        yellow + S + C/O + asteroids + sf & 0.264 & 0.285 & 0.206\\
        \hline
    \end{tabular}
    \caption{Probabilities of observing our values of the macroscopic variables for various combinations of habitability conditions mentioned in the text. For brevity, we only include $\beta$, $\gamma$ and $\kappa$ as the constants that are significantly influenced by the inclusion of star formation efficiency in our calculations.}
    \label{probs_table3}
\end{table*}

\noindent{\bf Most habitability combinations are incompatible with the multiverse\\}
To test the compatibility of each habitability condition including star formation efficiency and varying galactic density, we run through combinations of the various conditions outlined above. For expediency due to computational limitations, we do not consider all combinatorial possibilities, but restrict our attention to those combinations where a maximum of 8 habitability conditions above are chosen to be active, leading to a total of 34,074 combinations. In Fig. \ref{minprobfig} below, we display a histogram of the minimum probability of observing the 6 macroscopic constants and 3 local variables for each combination. From here it can be seen that the vast majority of these habitability combinations are incompatible with the multiverse hypothesis, since under these assumptions the probability of observing the value of at least one of our constants is exceedingly small. Among the possibilities we consider, the median minimum probability $p_\text{min}=7\times10^{-6}$, only $14\%$ of combinations have $p_\text{min}>.01$, and only $1.4\%$ of combinations have $p_\text{min}>.1$. This demonstrates that compatibility with the multiverse is the extreme exception, rather than the rule.\\

\begin{figure*}
\begin{center}
\includegraphics[width=.6\textwidth]{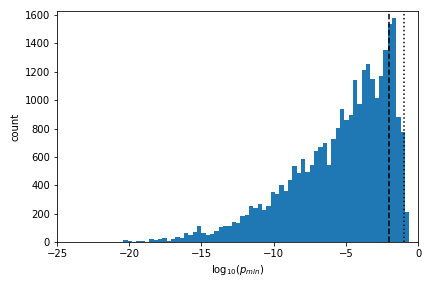}
\caption{The minimum probability of observing the 6 macroscopic constants and 3 local variables mentioned above for 34k combinations of habitability conditions. The dotted and dashed lines correspond to $p_\text{min}=.1$ and $.01$, respectively. The vast majority of combinations have exceedingly small probabilities, and so are incompatible with the multiverse.}
\label{minprobfig}
\end{center}
\end{figure*}

\begin{table*}
    \centering
    \begin{tabular}{|c|c|c|c|c|}
        \hline
        scenario & $W$ & $N(p_\text{min}>.1)$ & best condition & $\max(p_{min})$ \\
        \hline

        SI + lowE + SN & $\kappa^{1/2}\,\gamma^{1/4}$ & 32 & TL time C/O IDP sf & 0.186 \\
        SI + lowE & $\kappa^{1/2}\,\gamma$ & 15 & yellow area C/O lightning sf & 0.15\\
        SI + highE + SN & $\kappa^{7/8}\,\gamma^{1/4}$ & 19 & yellow TL Mg/Si moneta sf & 0.255\\
        SI + highE & $\kappa^{7/8}\,\gamma$ & 10 & yellow TL Mg/Si moneta sf & 0.229\\
        supA + lowE + SN & $\kappa^{7/8}\,\gamma^{3/4}$ & 13 & yellow TL Mg/Si moneta sf & 0.254\\
        supA + lowE & $\kappa^{7/8}\,\gamma^{3/2}$ & 10 & yellow area C/O comets sf & 0.16\\
        supA + highE + SN & $\kappa^{5/4}\,\gamma^{3/4}$ & 6 & yellow TL Mg/Si moneta sf & 0.213\\        
        supA + highE & $\kappa^{5/4}\,\gamma^{3/2}$ & 7 & yellow area C/O moneta sf & 0.188\\ 
        FO + lowE + SN & $\kappa^{5/4}\,\gamma^{5/4}$ & 7 & yellow area C/O moneta sf & 0.218\\
        FO + lowE & $\kappa^{5/4}\,\gamma^{2}$ & 1 & yellow area C/O moneta sf & 0.135\\
        FO + highE + SN & $\kappa^{13/8}\,\gamma^{5/4}$ & 0 & temp area Mg/Si XUV sf & 0.0585\\
        FO + highE & $\kappa^{13/8}\,\gamma^{2}$ & 0 & temp area Mg/Si XUV sf & 0.0821\\
        \hline
    \end{tabular}
    \caption{Effects of the various cosmology scenarios discussed in Table \ref{cosmo_table}. Included are the dependence of the induced weight $W$ on $\gamma$ and $\kappa$, the number of habitability conditions with minimum probability greater than .1 when coupled with the given cosmology scenario (out of 4,500), and the combination of habitability conditions yielding the highest minimum probability, along with its corresponding value.}
    \label{cosmo_probs_table}
\end{table*}
 
\noindent{\bf Theories with small dependence on $\kappa$ are favored\\}
Above, we restricted our attention to a particular cosmology scenario, {\bf SI lowE SN}, which dictated the power-law form of the weighting function for $\gamma$ and $\kappa$. Here, we scan over all choices in Table \ref{cosmo_table} to see how these choices affect which habitability conditions are viable. Due to computational limitations, for this full suite we restrict our combinations to a maximum of 5 habitability conditions at a time, rather than the maximum of 8 used when only a single cosmology scenario was considered. This leads to 4,500 combinations for each cosmology scenario, or 54,000 total.

With such a large number of combinations, it becomes quite cumbersome to narrativize and communicate results. One aspect that can be looked at is the number of combinations with $p_\text{min}$ above some threshold value as a function of cosmology scenario, displayed in Table \ref{cosmo_probs_table} for the value of .1. From this, it can be seen that the baseline scenario considered before {\bf SI lowE SN} contains the greatest number of habitability condition combinations that are compatible with the multiverse. Scenarios involving freeze-out dark matter fare poorest, and no scenario with both freeze-out dark matter and high energy baryogenesis contains a combination with $p_\text{min}>.1$. We also display the combination of habitability conditions yielding the highest $p_\text{min}$, along with its corresponding value. This shows that even for freeze-out scenarios, the probabilities aren't too far below .1, and so we cannot make any strong statements about a particular cosmology scenario being highly disfavored until we know more about the nature of habitability. One should not read too much into the particulars of the best conditions, since in each scenario there are other combinations with probabilities that are not too much smaller.

\begin{figure*}
\begin{center}
\includegraphics[height=.6\textwidth]{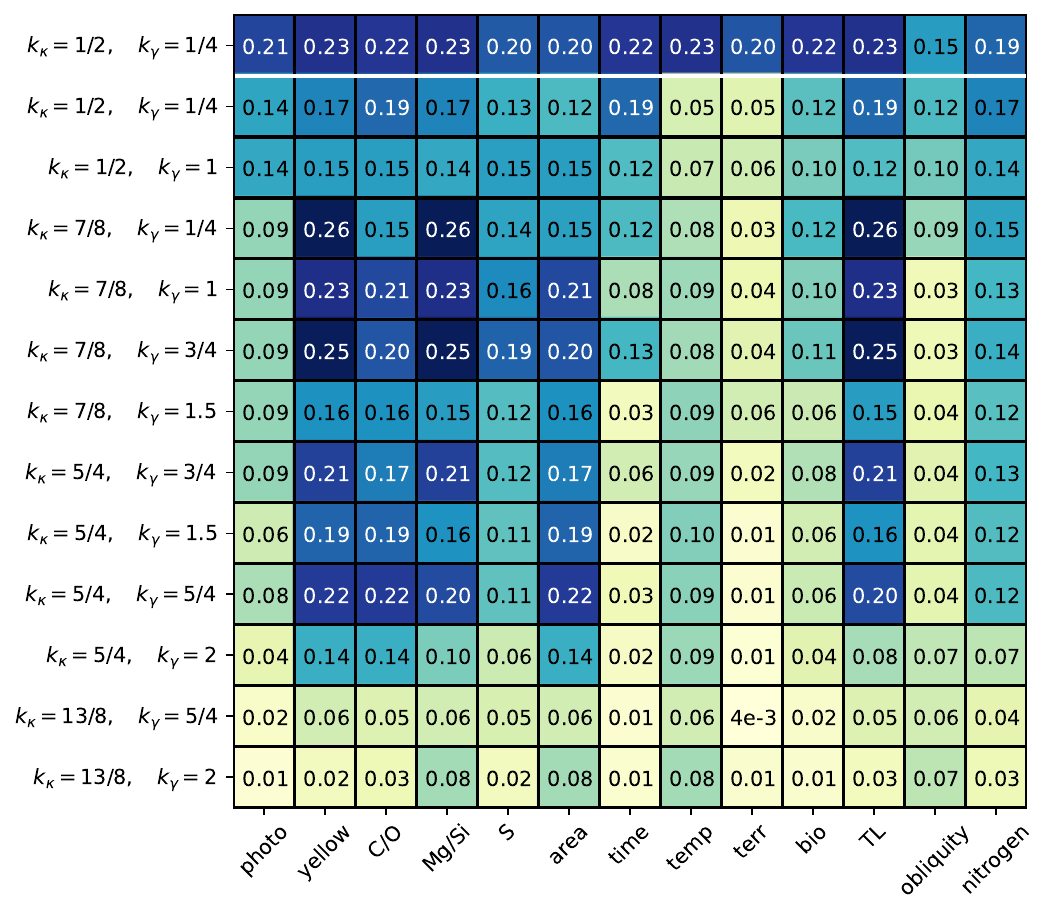}
\caption{Best value of the minimum probability for each cosmology scenario-habitability condition pair mentioned in the text. For each element in this chart, the minimum probability of the best combination of habitability conditions of all those which include the one mentioned is displayed. The top row includes combinations of up to 8 habitability conditions; the rest are capped at 5.}
\label{kkvsH}
\end{center}
\end{figure*}

To capture more information about the interaction between cosmology scenario and habitability conditions, in Fig. \ref{kkvsH} we display a matrix of combinations between the two. Here, the value displayed is the minimum probability $p_\text{min}$ for the best combination of habitability conditions containing the cosmology scenario indicated in that row and habitability condition indicated in that column. Each column corresponds to one of the 13 habitability conditions mentioned above, not including the origin of life scenarios. From here we can see that some cosmology scenarios are much more robustly compatible with the multiverse than others, and that the compatibility depends on the habitability conditions chosen. Broadly, those scenarios with lower $\kappa$ dependence are favored, echoing the results in Table \ref{cosmo_probs_table}. We can also see that some habitability conditions are nearly universally disfavored, including the {\bf temp}, {\bf terr}, and {\bf obliquity} conditions. The top row compares the baseline cosmology scenario with a maximum of 8 conditions to the rest of the rows, which contain a maximum of 5; the probabilities are higher for this top row compared to the second, as in each case there are more opportunities for a confluence of factors to allow any given habitability condition to ``piggyback" on some other combination. In these cases, the presence or absence of that habitability condition does not meaningfully alter the resulting probabilities. For cases with too many habitability conditions, the uncertainties in these reported probabilities start to become high, as the supports are lowered due to multiple thresholds winnowing down the random sample used to compute these values. For a single habitability condition, median support size for the sampling we use is $10^6$, which decreases steadily and dwindles to 46,000 for combinations with 8 conditions.\\\\\\\\\\\\



\begin{figure*}
\begin{center}
\includegraphics[height=.6\textwidth]{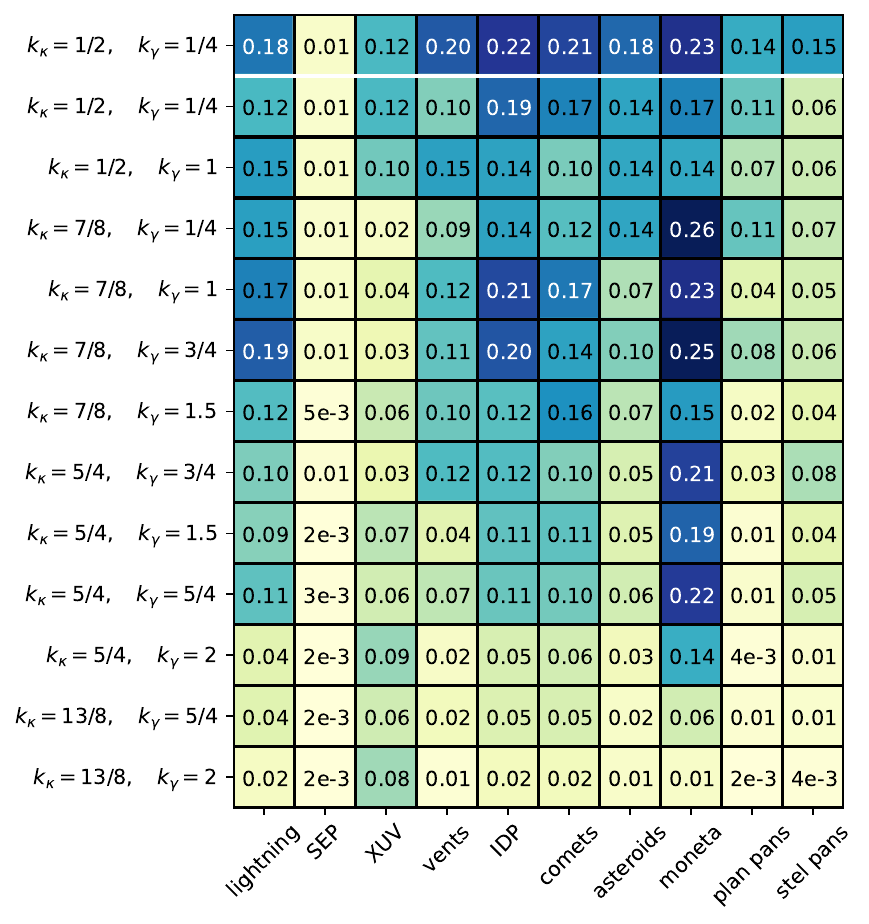}
\caption{Same as Fig. \ref{kkvsH} above, comparing against origin of life scenarios.}
\label{kkvsOOL}
\end{center}
\end{figure*}

\noindent{\bf Some origin of life theories are universally disfavored\\}
This combination analysis is repeated for the origin of life scenarios we consider in Fig. \ref{kkvsOOL}. We again see an overall trend that cosmology scenarios with steeper dependence on $\kappa$ are less favored. Additionally, we can see that some origin of life scenarios are nearly universally disfavored, such as {\bf SEP}, and to a lesser extent the {\bf XUV} and {\bf panspermia} scenarios.\\

\noindent{\bf Flexible GUTs are disfavored\\}
As discussed above, GUT theories can significantly alter the induced weight factor, as they enforce a constraint in parameter space. Though rigid GUTs (without WIMP dark matter) do not alter the weight factor, flexible GUTs introduce an extreme dependence on $\alpha$ and $\gamma$, as displayed in Table \ref{micro_table}. When including this factor for the {\bf yellow S C/O asteroids sf} condition, the minimum probability drops from .206 to $8\times10^{-13}$. This drop in probabilities holds more generally across different habitability conditions.

If dark matter is taken to be a WIMP, the induced weight has the form given in equation \ref{gutwimp}. To explore the effects of this we can use the {\bf FO lowE SN} scenario and the {\bf yellow area C/O moneta sf} condition, which yields the highest $p_\text{min}$ for a range of freeze-out dark matter scenarios. When including the GUT induced weight, the probabilities are altered by at most $15\%$, and so it is possible for rigid GUT theories to be compatible with the multiverse, both in WIMP and non-WIMP dark matter scenarios.\\

\noindent{\bf Pessimistic bounds for disruptions are disfavored\\}
Lastly, we include the galactic effects mentioned in section \ref{galhab}, namely stellar encounters, supernova explosions, and active galactic nuclei. These factors will adversely impact habitability for high galactic density, though the importance of each of these is currently debated. To determine the interplay of each of these with our multiverse calculations, we adopt the forms discussed above, all of which include a free parameter that can be cast as the fraction of systems that are unaffected by that type of disruption, $p_\text{survive}$. Though this number is usually considered to be quite close to 1 in our universe, it scales exponentially with constants in all three cases, and so can have a large impact on the habitability of other universes. We display how this affects probabilities for three different choices of $p_\text{survive}=.99,.9,.5$ in Table \ref{disrupt_table}.

\begin{table*}
    \centering
    \begin{tabular}{|c|c|c|c|c|c|c|}
        \hline
        condition & $P(\alpha)$ & $P(\beta)$ & $P(\gamma)$ & $P(\delta_u)$ & $P(\delta_d)$ & $P(\kappa)$\\
        \hline
        $p_\text{survive}=1$ &&&&&&\\
        yellow S C/O asteroids & 0.408 & 0.264 & 0.285 & 0.425 & 0.485 & 0.206\\
        \hline
        $p_\text{survive}=.99$ &&&&&&\\
        stellar encounters & 0.426 & 0.24 & 0.329 & 0.264 & 0.316 & 0.319\\
        SN & 0.432 & 0.495 & 0.484 & 0.425 & 0.364 & 0.0237\\
        AGN & 0.337 & 0.358 & 0.359 & 0.31 & 0.214 & 0.358\\
        \hline
        $p_\text{survive}=.9$ &&&&&&\\
        stellar encounters & 0.168 & 0.105 & 0.224 & 0.174 & 0.0196 & 0.364\\
        SN & 0.467 & 0.33 & 0.399 & 0.475 & 0.262 & 0.00434\\
        AGN & 0.174 & 0.353 & 0.407 & 0.175 & 0.0795 & 0.405\\
        \hline
        $p_\text{survive}=.5$ &&&&&&\\
        stellar encounters & 0.0398 & 0.112 & 0.101 & 0.0399 & 0.0118 & 0.181 \\
        SN & 0.213 & 0.14 & 0.193 & 0.22 & 0.15 & 0.00601\\
        AGN & 0.062 & 0.261 & 0.38 & 0.0625 & 0.0528 & 0.134\\
        \hline
        \end{tabular}
    \caption{Probability of the various macroscopic constants for differing galactic habitability effects. All probabilities reported use the {\bf yellow S C/O asteroids} condition and the baseline cosmology scenario {\bf SI lowE SN}. The strength of these effects are parameterized by $p_\text{survive}$, the fraction of planetary systems which are unaffected by the effect in question. The top row is the null hypothesis, in which the effect is completely negligible.}
    \label{disrupt_table}
\end{table*}

From this table, we can see that even for $p_\text{survive}=.99$, including these disruption scenarios induces some tension in the multiverse setting, particularly for the supernovae. For $p_\text{survive}=.9$, all three disruption scenarios are disfavored for at least one probability, though in general the AGN scenario is least affected. Though the reported results are only for a single habitability condition, this illustrates a more general conclusion, that in the multiverse setting, we should not expect our universe to be on the ``hairy edge'' of habitability when it comes to galactic disruption events. Given the tendency for these galactic effects to favor small $\kappa$, we checked that adding the {\bf terr} habitability condition, which disfavors small $\kappa$, does not affect these conclusions.\\

\noindent{\bf Conclusions \\}
We've shown here that within the multiverse setting, the probabilities of observing our values of the macroscopic and local variables depend sensitively on the assumptions we make about fundamental physics and galactic habitability. In conjunction with previous works in this series exploring the sensitivity to other habitability assumptions, this establishes that it is indeed possible to make concrete testable predictions within the multiverse setting. Testing these predictions in future experiments can allow us to gain quite strong evidence for whether the multiverse is true or not.

Our predictions in this paper have tended not to be blanket statements unequivocally declaring the truth of a particular theory of fundamental physics or habitability. Rather, they are more nuanced, finding that particular scenarios are (dis)allowed, subject to certain conditions. As we gain more information on fundamental theories and habitability conditions, we will be able to iterate on this process, allowing for more succinct but no less exact predictions. Lest a critic object, we hasten to note that in the course of our analysis we did make several robust findings, for instance that freeze-out dark matter in conjunction with high energy baryogenesis is disfavored, several origin of life scenarios are disfavored, flexible GUTs are highly disfavored, and that galactic effects should not play a large role in limiting habitability in our universe.

Our analysis throughout these papers can in no sense be considered complete or final. We believe at this point that we have included the most important effects dictating habitability as we currently understand it, but there is no guarantee that some new discoveries will not introduce additional factors that may alter even our broad brush conclusions. In this eventuality we may include these considerations and recompute our probabilities within our existing framework as they arise (code is available here: \url{https://github.com/mccsandora/Multiverse-Habitability-Handler}). Additionally, many of our estimates have been decidedly back of the envelope, and future work adding a higher level of detail to these is needed to fully account for their effects. Even in the event that many of our current predictions need substantial revision, we take the importance of this work to be the framework developed which allows us to provide an answer to the fundamental question of whether or not other universes exist.



\smallskip

\bibliographystyle{apalike}
\bibliography{yellowbib}

\begin{thebibliography}{}

\bibitem[Adams et~al., 2015]{ACBplanets}
Adams, F.~C., Coppess, K.~R., and Bloch, A.~M. (2015).
\newblock Planets in other universes: habitability constraints on density fluctuations and galactic structure.
\newblock {\em Journal of Cosmology and Astroparticle Physics}, 2015(09):030.

\bibitem[Affleck and Dine, 1985]{affleck1985new}
Affleck, I. and Dine, M. (1985).
\newblock A new mechanism for baryogenesis.
\newblock {\em Nuclear Physics B}, 249(2):361--380.

\bibitem[Arvanitaki et~al., 2010]{arvanitaki2010string}
Arvanitaki, A., Dimopoulos, S., Dubovsky, S., Kaloper, N., and March-Russell, J. (2010).
\newblock String axiverse.
\newblock {\em Physical Review D—Particles, Fields, Gravitation, and Cosmology}, 81(12):123530.

\bibitem[Balbi and Tombesi, 2017]{balbi2017habitability}
Balbi, A. and Tombesi, F. (2017).
\newblock The habitability of the milky way during the active phase of its central supermassive black hole.
\newblock {\em Scientific Reports}, 7(1):16626.

\bibitem[Barbieri et~al., 2000]{barbieri2000baryogenesis}
Barbieri, R., Creminelli, P., Strumia, A., and Tetradis, N. (2000).
\newblock Baryogenesis through leptogenesis.
\newblock {\em Nuclear Physics B}, 575(1-2):61--77.

\bibitem[Barnes et~al., 2018]{barnes2018galaxy}
Barnes, L.~A., Elahi, P.~J., Salcido, J., Bower, R.~G., Lewis, G.~F., Theuns, T., Schaller, M., Crain, R.~A., and Schaye, J. (2018).
\newblock Galaxy formation efficiency and the multiverse explanation of the cosmological constant with eagle simulations.
\newblock {\em Monthly Notices of the Royal Astronomical Society}, 477(3):3727--3743.

\bibitem[Behroozi et~al., 2013]{behroozi2013average}
Behroozi, P.~S., Wechsler, R.~H., and Conroy, C. (2013).
\newblock The average star formation histories of galaxies in dark matter halos from z= 0--8.
\newblock {\em The Astrophysical Journal}, 770(1):57.

\bibitem[Bousso and Hall, 2013]{bousso2013comparable}
Bousso, R. and Hall, L. (2013).
\newblock Why comparable? a multiverse explanation of the dark matter-baryon coincidence.
\newblock {\em Physical Review D—Particles, Fields, Gravitation, and Cosmology}, 88(6):063503.

\bibitem[Bousso and Leichenauer, 2010]{bousso2010predictions}
Bousso, R. and Leichenauer, S. (2010).
\newblock Predictions from star formation in the multiverse.
\newblock {\em Physical Review D—Particles, Fields, Gravitation, and Cosmology}, 81(6):063524.

\bibitem[Buchner, 2024]{buchner2024impediments}
Buchner, J. (2024).
\newblock Impediments to the cosmic growth of galaxies: The outflow budget from star formation and active galactic nuclei.
\newblock {\em Astronomy \& Astrophysics}, 689:L2.

\bibitem[Carr and Rees, 1979]{carr1979anthropic}
Carr, B.~J. and Rees, M.~J. (1979).
\newblock The anthropic principle and the structure of the physical world.
\newblock {\em Nature}, 278(5705):605--612.

\bibitem[Clavelli and White~III, 2006]{clavelli2006problems}
Clavelli, L. and White~III, R. (2006).
\newblock Problems in a weakless universe.
\newblock {\em arXiv preprint hep-ph/0609050}.

\bibitem[Coleman and Weinberg, 1973]{coleman1973radiative}
Coleman, S. and Weinberg, E. (1973).
\newblock Radiative corrections as the origin of spontaneous symmetry breaking.
\newblock {\em Physical Review D}, 7(6):1888.

\bibitem[Davidson et~al., 2008]{davidson2008leptogenesis}
Davidson, S., Nardi, E., and Nir, Y. (2008).
\newblock Leptogenesis.
\newblock {\em Physics Reports}, 466(4-5):105--177.

\bibitem[De~Simone et~al., 2008]{scale}
De~Simone, A., Guth, A.~H., Salem, M.~P., and Vilenkin, A. (2008).
\newblock Predicting the cosmological constant with the scale-factor cutoff measure.
\newblock {\em Phys.~Rev.~D}, 78(6).

\bibitem[Dekel and Silk, 1986]{dekel1986origin}
Dekel, A. and Silk, J. (1986).
\newblock The origin of dwarf galaxies, cold dark matter, and biased galaxy formation.
\newblock {\em Astrophysical Journal, Part 1 (ISSN 0004-637X), vol. 303, April 1, 1986, p. 39-55.}, 303:39--55.

\bibitem[{Donoghue} et~al., 2006]{leptonland}
{Donoghue}, J.~F., {Dutta}, K., and {Ross}, A. (2006).
\newblock {Quark and lepton masses and mixing in the landscape}.
\newblock {\em Phys.~Rev.~D}, 73(11):113002.

\bibitem[D’Amico et~al., 2019]{d2019direct}
D’Amico, G., Strumia, A., Urbano, A., and Xue, W. (2019).
\newblock Direct anthropic bound on the weak scale from supernov{\ae} explosions.
\newblock {\em Physical Review D}, 100(8):083013.

\bibitem[Feldstein et~al., 2005]{feldstein2005density}
Feldstein, B., Hall, L.~J., and Watari, T. (2005).
\newblock Density perturbations and the cosmological constant from inflationary landscapes.
\newblock {\em Physical Review D—Particles, Fields, Gravitation, and Cosmology}, 72(12):123506.

\bibitem[Feng, 2023]{feng2023wimp}
Feng, J.~L. (2023).
\newblock The wimp paradigm: Theme and variations.
\newblock {\em SciPost Physics Lecture Notes}, page 071.

\bibitem[Fields et~al., 2020]{fields2020supernova}
Fields, B.~D., Melott, A.~L., Ellis, J., Ertel, A.~F., Fry, B.~J., Lieberman, B.~S., Liu, Z., Miller, J.~A., and Thomas, B.~C. (2020).
\newblock Supernova triggers for end-devonian extinctions.
\newblock {\em Proceedings of the National Academy of Sciences}, 117(35):21008--21010.

\bibitem[Freivogel, 2010]{freivogel2010anthropic}
Freivogel, B. (2010).
\newblock Anthropic explanation of the dark matter abundance.
\newblock {\em Journal of Cosmology and Astroparticle Physics}, 2010(03):021.

\bibitem[{Freivogel}, 2011]{Fmeasure}
{Freivogel}, B. (2011).
\newblock {Making predictions in the multiverse}.
\newblock {\em Classical and Quantum Gravity}, 28(20):204007.

\bibitem[Froggatt and Nielsen, 1996]{froggatt1996standard}
Froggatt, C. and Nielsen, H.~B. (1996).
\newblock Standard model criticality prediction top mass 173$\pm$5 gev and higgs mass 135$\pm$9 gev.
\newblock {\em Physics Letters B}, 368(1-2):96--102.

\bibitem[Garny et~al., 2016]{garny2016planckian}
Garny, M., Sandora, M., and Sloth, M.~S. (2016).
\newblock Planckian interacting massive particles as dark matter.
\newblock {\em Physical review letters}, 116(10):101302.

\bibitem[{Garriga} et~al., 2000]{GLV}
{Garriga}, J., {Livio}, M., and {Vilenkin}, A. (2000).
\newblock {Cosmological constant and the time of its dominance}.
\newblock {\em Phys.~Rev.~D}, 61(2):023503.

\bibitem[Garriga and Vilenkin, 2006]{hep-th/0508005}
Garriga, J. and Vilenkin, A. (2006).
\newblock Anthropic prediction for $\lambda$ and the q catastrophe.
\newblock {\em Progress of Theoretical Physics Supplement}, 163:245--257.

\bibitem[Gonzalez, 2005]{gonzalez2005habitable}
Gonzalez, G. (2005).
\newblock Habitable zones in the universe.
\newblock {\em Origins of life and evolution of biospheres}, 35(6):555--606.

\bibitem[Gonzalez et~al., 2001]{Gonzalez:2001hh}
Gonzalez, G., Brownlee, D., and Ward, P. (2001).
\newblock The galactic habitable zone: galactic chemical evolution.
\newblock {\em Icarus}, 152(1):185--200.

\bibitem[Green and Kavanagh, 2021]{green2021primordial}
Green, A.~M. and Kavanagh, B.~J. (2021).
\newblock Primordial black holes as a dark matter candidate.
\newblock {\em Journal of Physics G: Nuclear and Particle Physics}, 48(4):043001.

\bibitem[Hall et~al., 2010]{hall2010freeze}
Hall, L.~J., Jedamzik, K., March-Russell, J., and West, S.~M. (2010).
\newblock Freeze-in production of fimp dark matter.
\newblock {\em Journal of High Energy Physics}, 2010(3):1--33.

\bibitem[Hall and Nomura, 2008]{hall2008evidence}
Hall, L.~J. and Nomura, Y. (2008).
\newblock Evidence for the multiverse in the standard model and beyond.
\newblock {\em Physical Review D}, 78(3):035001.

\bibitem[Hall et~al., 2014]{hall2014weak}
Hall, L.~J., Pinner, D., and Ruderman, J.~T. (2014).
\newblock The weak scale from bbn.
\newblock {\em Journal of High Energy Physics}, 2014(12):1--29.

\bibitem[H{\"a}ring and Rix, 2004]{haring2004black}
H{\"a}ring, N. and Rix, H.-W. (2004).
\newblock On the black hole mass-bulge mass relation.
\newblock {\em The Astrophysical Journal}, 604(2):L89.

\bibitem[Harnik et~al., 2006]{harnik2006universe}
Harnik, R., Kribs, G.~D., and Perez, G. (2006).
\newblock A universe without weak interactions.
\newblock {\em Physical Review D—Particles, Fields, Gravitation, and Cosmology}, 74(3):035006.

\bibitem[Hertzberg, 2017]{hertzberg2017correlation}
Hertzberg, M.~P. (2017).
\newblock A correlation between the higgs mass and dark matter.
\newblock {\em Advances in High Energy Physics}, 2017(1):6295927.

\bibitem[Hiller et~al., 2024]{hiller2024vacuum}
Hiller, G., H{\"o}hne, T., Litim, D.~F., and Steudtner, T. (2024).
\newblock Vacuum stability in the standard model and beyond.
\newblock {\em Physical Review D}, 110(11):115017.

\bibitem[Isidori et~al., 2001]{isidori2001metastability}
Isidori, G., Ridolfi, G., and Strumia, A. (2001).
\newblock On the metastability of the standard model vacuum.
\newblock {\em Nuclear Physics B}, 609(3):387--409.

\bibitem[Janka, 2012]{janka2012explosion}
Janka, H.-T. (2012).
\newblock Explosion mechanisms of core-collapse supernovae.
\newblock {\em Annual Review of Nuclear and Particle Science}, 62(1):407--451.

\bibitem[Kaib and Raymond, 2025]{kaib2025influence}
Kaib, N.~A. and Raymond, S.~N. (2025).
\newblock The influence of passing field stars on the solar system’s dynamical future.
\newblock {\em Icarus}, page 116632.

\bibitem[King, 2003]{king2003black}
King, A. (2003).
\newblock Black holes, galaxy formation, and the mbh-$\sigma$ relation.
\newblock {\em The Astrophysical Journal}, 596(1):L27.

\bibitem[Kokaia and Davies, 2019]{kokaia2019stellar}
Kokaia, G. and Davies, M.~B. (2019).
\newblock Stellar encounters with giant molecular clouds.
\newblock {\em Monthly Notices of the Royal Astronomical Society}, 489(4):5165--5180.

\bibitem[Kolb and Turner, 1990]{Kolb:1990vq}
Kolb, E.~W. and Turner, M.~S. (1990).
\newblock {\em {The Early Universe}}, volume~69.
\newblock Taylor and Francis.

\bibitem[Li and Adams, 2015]{li2015cross}
Li, G. and Adams, F.~C. (2015).
\newblock Cross-sections for planetary systems interacting with passing stars and binaries.
\newblock {\em Monthly Notices of the Royal Astronomical Society}, 448(1):344--363.

\bibitem[Linde, 1985]{linde1985new}
Linde, A.~D. (1985).
\newblock A new mechanism of baryogeneses and the inflationary universe.
\newblock {\em Physics Letters B}, 160(4-5):243--248.

\bibitem[Lingam et~al., 2019]{lingam2019active}
Lingam, M., Ginsburg, I., and Bialy, S. (2019).
\newblock Active galactic nuclei: boon or bane for biota?
\newblock {\em The Astrophysical Journal}, 877(1):62.

\bibitem[Marsh, 2016]{marsh2016axion}
Marsh, D.~J. (2016).
\newblock Axion cosmology.
\newblock {\em Physics Reports}, 643:1--79.

\bibitem[Mirizzi et~al., 2016]{mirizzi2016supernova}
Mirizzi, A., Tamborra, I., Janka, H.-T., Saviano, N., Scholberg, K., Bollig, R., H{\"u}depohl, L., and Chakraborty, S. (2016).
\newblock Supernova neutrinos: production, oscillations and detection.
\newblock {\em La Rivista del Nuovo Cimento}, 39:1--112.

\bibitem[Morrissey and Ramsey-Musolf, 2012]{morrissey2012electroweak}
Morrissey, D.~E. and Ramsey-Musolf, M.~J. (2012).
\newblock Electroweak baryogenesis.
\newblock {\em New Journal of Physics}, 14(12):125003.

\bibitem[Mukhanov, 2005]{mukhanov2005physical}
Mukhanov, V.~F. (2005).
\newblock {\em Physical foundations of cosmology}.
\newblock Cambridge university press.

\bibitem[Nanopoulos, 1980]{nanopoulos1980bounds}
Nanopoulos, D. (1980).
\newblock Bounds on the baryon/photon ratio due to our existence.
\newblock {\em Physics Letters B}, 91(1):67--71.

\bibitem[Niedermann and Sloth, 2023]{niedermann2023new}
Niedermann, F. and Sloth, M.~S. (2023).
\newblock New early dark energy as a solution to the $ h\_0 $ and $ s\_8 $ tensions.
\newblock {\em arXiv preprint arXiv:2307.03481}.

\bibitem[Oh et~al., 2022]{oh2022fate}
Oh, B.~K., Peacock, J.~A., Khochfar, S., and Smith, B.~D. (2022).
\newblock The fate of baryons in counterfactual universes.
\newblock {\em Monthly Notices of the Royal Astronomical Society}, 517(1):59--75.

\bibitem[Particle Data~Group\ et~al., 2022]{particle2022review}
Particle Data~Group\, and~Workman, R., Burkert, V., Crede, V., Klempt, E., Thoma, U., Tiator, L., Agashe, K., Aielli, G., Allanach, B., et~al. (2022).
\newblock Review of particle physics.
\newblock {\em Progress of theoretical and experimental physics}, 2022(8):083C01.

\bibitem[Peskin, 2018]{peskin2018introduction}
Peskin, M.~E. (2018).
\newblock {\em An Introduction to quantum field theory}.
\newblock CRC press.

\bibitem[Piran et~al., 2016]{1508.01034}
Piran, T., Jimenez, R., Cuesta, A.~J., Simpson, F., and Verde, L. (2016).
\newblock Cosmic explosions, life in the universe, and the cosmological constant.
\newblock {\em Physical review letters}, 116(8):081301.

\bibitem[Pogosian and Vilenkin, 2007]{astro-ph/0611573}
Pogosian, L. and Vilenkin, A. (2007).
\newblock Anthropic predictions for vacuum energy and neutrino masses in the light of wmap-3.
\newblock {\em Journal of Cosmology and Astroparticle Physics}, 2007(01):025.

\bibitem[Salcido et~al., 2020]{salcido2020feedback}
Salcido, J., Bower, R.~G., and Theuns, T. (2020).
\newblock How feedback shapes galaxies: an analytic model.
\newblock {\em Monthly Notices of the Royal Astronomical Society}, 491(4):5083--5100.

\bibitem[Sandora, 2019a]{mc4}
Sandora, M. (2019a).
\newblock Multiverse predictions for habitability: Fraction of life that develops intelligence.
\newblock {\em Universe}, 5(7):175.

\bibitem[Sandora, 2019b]{mc3}
Sandora, M. (2019b).
\newblock Multiverse predictions for habitability: Fraction of planets that develop life.
\newblock {\em Universe}, 5(7):171.

\bibitem[Sandora, 2019c]{mc2}
Sandora, M. (2019c).
\newblock Multiverse predictions for habitability: Number of potentially habitable planets.
\newblock {\em Universe}, 5(6):157.

\bibitem[Sandora, 2019d]{mc1}
Sandora, M. (2019d).
\newblock Multiverse predictions for habitability: The number of stars and their properties.
\newblock {\em Universe}, 5(6):149.

\bibitem[Sandora et~al., 2022a]{mc6}
Sandora, M., Airapetian, V., Barnes, L., and Lewis, G.~F. (2022a).
\newblock Multiverse predictions for habitability: Planetary characteristics.
\newblock {\em Universe}, 9(1):2.

\bibitem[Sandora et~al., 2022b]{mc5}
Sandora, M., Airapetian, V., Barnes, L., Lewis, G.~F., and P{\'e}rez-Rodr{\'\i}guez, I. (2022b).
\newblock Multiverse predictions for habitability: Element abundances.
\newblock {\em Universe}, 8(12):651.

\bibitem[Sandora et~al., 2023]{mc8}
Sandora, M., Airapetian, V., Barnes, L., Lewis, G.~F., and P{\'e}rez-Rodr{\'\i}guez, I. (2023).
\newblock Multiverse predictions for habitability: Origin of life scenarios.
\newblock {\em Universe}, 9(1):42.

\bibitem[Silk and Mamon, 2012]{silk2012current}
Silk, J. and Mamon, G.~A. (2012).
\newblock The current status of galaxy formation.
\newblock {\em Research in Astronomy and Astrophysics}, 12(8):917.

\bibitem[Sorini et~al., 2024]{sorini2024impact}
Sorini, D., Peacock, J.~A., and Lombriser, L. (2024).
\newblock The impact of the cosmological constant on past and future star formation.
\newblock {\em Monthly Notices of the Royal Astronomical Society}, 535(2):1449--1474.

\bibitem[Starobinsky, 1982]{starobinsky1982dynamics}
Starobinsky, A.~A. (1982).
\newblock Dynamics of phase transition in the new inflationary universe scenario and generation of perturbations.
\newblock {\em Physics Letters B}, 117(3-4):175--178.

\bibitem[Svrcek and Witten, 2006]{svrcek2006axions}
Svrcek, P. and Witten, E. (2006).
\newblock Axions in string theory.
\newblock {\em Journal of High Energy Physics}, 2006(06):051.

\bibitem[Tegmark et~al., 2006]{matters}
Tegmark, M., Aguirre, A., Rees, M.~J., and Wilczek, F. (2006).
\newblock Dimensionless constants, cosmology, and other dark matters.
\newblock {\em Physical Review D}, 73(2):023505.

\bibitem[{Tegmark} and {Rees}, 1998]{starstar}
{Tegmark}, M. and {Rees}, M.~J. (1998).
\newblock {Why Is the Cosmic Microwave Background Fluctuation Level 10$^{-5}$?}
\newblock {\em Astrophysical Journal}, 499:526--532.

\bibitem[Thomas et~al., 2016]{thomas2016terrestrial}
Thomas, B.~C., Engler, E., Kachelrie{\ss}, M., Melott, A., Overholt, A., and Semikoz, D. (2016).
\newblock Terrestrial effects of nearby supernovae in the early pleistocene.
\newblock {\em The Astrophysical journal letters}, 826(1):L3.

\bibitem[Totani et~al., 2019]{totani2019lethal}
Totani, T., Omiya, H., Sudoh, T., Kobayashi, M.~A., and Nagashima, M. (2019).
\newblock Lethal radiation from nearby supernovae helps explain the small cosmological constant.
\newblock {\em Astrobiology}, 19(1):126--131.

\bibitem[Trodden, 1999]{trodden1999electroweak}
Trodden, M. (1999).
\newblock Electroweak baryogenesis.
\newblock {\em Reviews of Modern Physics}, 71(5):1463.

\bibitem[{Vilenkin}, 1995]{mediocre}
{Vilenkin}, A. (1995).
\newblock {Predictions from Quantum Cosmology}.
\newblock {\em Physical Review Letters}, 74:846--849.

\bibitem[{Weinberg}, 1987]{cc}
{Weinberg}, S. (1987).
\newblock {Anthropic bound on the cosmological constant}.
\newblock {\em Physical Review Letters}, 59:2607--2610.

\bibitem[Whitmire, 2020]{whitmire2020habitability}
Whitmire, D.~P. (2020).
\newblock The habitability of large elliptical galaxies.
\newblock {\em Monthly Notices of the Royal Astronomical Society}, 494(2):3048--3052.

\bibitem[Zee, 2010]{zee2010quantum}
Zee, A. (2010).
\newblock {\em Quantum field theory in a nutshell}, volume~7.
\newblock Princeton university press.

\end{thebibliography}

\end{document}